\documentclass[aps,prd,reprint,superscriptaddress,nofootinbib,floatfix]{revtex4-2}

\usepackage{xcolor}

\usepackage[T1]{fontenc}
\usepackage{amsmath,amssymb}
\usepackage{bm}
\usepackage{booktabs}
\usepackage{graphicx}
\usepackage{float}
\usepackage[colorlinks=true,linkcolor=blue,citecolor=blue,urlcolor=blue]{hyperref}

\newcommand{\Msun}{M_\odot}
\newcommand{\eibi}{Eddington-inspired Born-Infeld~}
\newcommand{\RsunG}{\mathcal{R}_\odot}
\newcommand{\epshat}{\widehat{\epsilon}}
\newcommand{\phat}{\widehat{P}}
\newcommand{\Sigmahat}{\widehat{\Sigma}}

\begin{document}

\title{Tidal deformability of white dwarfs in Eddington-inspired Born-Infeld gravity}

\author{Santanu Pradhan}
\email{santanup25@iitk.ac.in}
\affiliation{Department of Physics, Indian Institute of Technology Kanpur, Kanpur 208016, India}
\author{Tapobrata Sarkar}
\email{tapo@iitk.ac.in}
\affiliation{Department of Physics, Indian Institute of Technology Kanpur, Kanpur 208016, India}

\begin{abstract}
We compute the static quadrupole Love number $k_2$ and the dimensionless tidal deformability $\Lambda$ of cold white dwarfs in Eddington-inspired Born-Infeld (EiBI) gravity. Matter couples to the physical metric while perturbation equations are solved in an auxiliary metric, so the physical tidal field near the surface depends on derivatives of the equation of state. This dependence is regular for a surface polytropic index $1<\Gamma_{\rm env}<3/2$, but not for the $\Gamma = 5/3$ Chandrasekhar equation of state. The mechanism, an algebraic map between the metric that carries perturbations and the metric that matter feels, is common to Palatini-type theories, so the same type of surface condition should apply to tidal calculations in this wider class. We use thermodynamically consistent envelopes that exactly reach zero enthalpy. At five masses from \(0.6\) to \(1.3\,\Msun\) and five sampled Born-Infeld couplings \(\kappa\), with $|\kappa|\le5\times10^6\,\mathrm{m^2}$, the EiBI-GR differences are approximately linear in $\kappa$. The largest absolute fractional shifts are \(3.89\times 10^{-5}\) in the stellar radius $R$, \(4.09\times 10^{-5}\) in $k_2$, and \(2.36\times 10^{-4}\) in $\Lambda$. 
Across nine envelope models, half the spread of each quantity relative to the baseline envelope is at most \(0.0186\) in $R$ and \(0.0857\) in \(k_2\),
but only \(3.02\times 10^{-5}\) in \(\Lambda\) as the changes in \(k_2\) and \(R^5\) largely cancel. 
A high-precision calculation shows that the envelope dependence of the EiBI-GR difference in \(\Lambda\) is small. Half the spread of this difference ranges from \(1.067\times 10^{-12}\) to \(6.208\times 10^{-11}\) over the 20 nonzero mass-coupling points.
\end{abstract}

\maketitle

\section{Introduction}
\label{sec:introduction}

An external tidal field deforms a star and induces a mass quadrupole moment. The static quadrupole Love number $k_2$ describes the response. The corresponding dimensionless tidal deformability is
\begin{equation}
 \Lambda=\frac{2k_2}{3C^5},
 \label{eq:tidal_def}
\end{equation}
where $C=GM/(Rc^2)$ is the compactness of a star with gravitational mass \(M\) and radius \(R\). Here \(G\) is Newton's gravitational constant and \(c\) is the speed of light. The response depends on how matter is distributed inside the star, as well as on the gravitational field equations. Thus $k_2$ and $\Lambda$ can distinguish stellar models that have similar mass-radius relations.

\eibi (EiBI) gravity is a natural theory in which to make this comparison. In its Palatini formulation, the spacetime metric and affine connection are varied independently \cite{banados2010}. Matter couples minimally to the physical metric $g_{\mu\nu}$, while the connection \(\Gamma^\alpha{}_{\mu\nu}\) is compatible with an auxiliary metric $q_{\mu\nu}$. The local stress-energy tensor algebraically relates $q_{\mu\nu}$ to $g_{\mu\nu}$ \cite{pani2012phenomenology,delsate2012}. With a vanishing cosmological constant, the vacuum equations reduce to general relativity (GR). Any change in the exterior tidal response must therefore enter through the stellar interior and the surface data matched to the Schwarzschild solution.
White dwarfs have small compactness and extended outer layers. In EiBI gravity, the perturbation equations take their simples forms in an auxiliary geometry. The physical tidal perturbation follows from the auxiliary one through an algebraic map that depends on the local pressure and energy density, so its radial derivative contains derivatives of these quantities. Such terms are absent in GR. Near the surface, where the pressure and sound speed go to zero, they need not vanish, and the physical and auxiliary tidal derivatives can then differ.

This sensitivity is controlled by
\begin{equation}
 \Sigma\equiv\frac{\epsilon+P}{c_s^2},
 \qquad c_s^2\equiv\frac{dP}{d\epsilon},
 \label{eq:sigma-definition}
\end{equation}
where \(\epsilon\) is the total energy density and \(c_s^2\) is the squared adiabatic sound speed (in units of \(c^2\)). Although \(P\), \(\epsilon\), and \(c_s^2\) vanish at the surface, their ratio need not have a regular radial derivative. The low-density equation of state fixes the limiting behavior of \(\Sigma\), which enters the metric deformation variable \(\chi\) defined below. Stopping at a small nonzero pressure bypasses this limit. It can leave the inferred tidal derivative dependent on the stopping pressure, even when the numerical endpoint is located accurately.

The standard zero-temperature Chandrasekhar equation of state (EOS) approaches \(P\propto\rho^{5/3}\) at low density, where the electrons are nonrelativistic. With this behavior, the physical tidal derivative does not have a regular limit at zero density. We retain the Chandrasekhar EOS in the degenerate core and replace only the low-density edge with an envelope whose pressure and energy density satisfy the first law of thermodynamics.

We then ask two questions. How does EiBI gravity change the structure and tidal response of a white dwarf of given mass? And how much do the predictions depend on the choice of surface model? We address these using mass-radius sequences, Love numbers and tidal deformabilities. 

Compact white dwarf binaries motivate this static response calculation. Eclipsing systems such as the 6.91-minute binary ZTF J153932.16+502738.8 provide measured orbital decay and potential sources for the Laser Interferometer Space Antenna (LISA) \cite{burdge2019}. Their tides need not be static: for a circular orbit with aligned spin, the dominant quadrupole forcing frequency in the rotating star is \(2(\Omega-\Omega_s)\), where \(\Omega\) and \(\Omega_s\) are the orbital and spin angular frequencies \cite{fuller2012}. Our static calculation gives an equilibrium response, without modeling synchronization, tidal heating, or detectable EiBI signals.

\subsection{Contribution of the Present Work}
\label{sec:contributions}

Relativistic EiBI studies have examined compact star equilibrium and radial stability, phase transition anomalies, and neutron star tidal observables \cite{sham2012,sham2013,sham2014,prasetyo2021}. Other EiBI applications have considered compact stars and white dwarf equilibrium structure as well as astrophysical constraints \cite{pani2012phenomenology,banerjee2017,avelino2012}. Another relevant comparison is the relativistic white dwarf tidal calculation in GR \cite{perot2022}. To the best of our knowledge, these studies do not calculate the static quadrupole Love number or dimensionless tidal deformability of a white dwarf in relativistic EiBI gravity. Such a calculation needs a regular physical surface, a surface value for matching to the exterior, and a test of how much the results depend on the low-density EOS. 

We use the standard EiBI structure and static tidal equations, Schwarzschild matching, and the Chandrasekhar EOS for the core. The tidal equation is solved for the logarithmic derivative \(y_F = rF'/F\), where \(F\) is the temporal tidal perturbation of the auxiliary metric. 
We integrate it from a regular center where \(y_F=\ell\), with  \(\ell\) being the multipole order to the zero-enthalpy surface \(h=0\) where \(h\) is the relativistic enthalpy
defined in sequel in Sec.~\ref{sec:gravity}. What is new is the treatment of this surface. We derive the conversion from \(y_F\) to the logarithmic derivative of
the physical metric perturbation with respect to the circumferential radius \(R_{\rm circ}\), and find when the two agree at the surface. 

The surface analysis in Eq.~\eqref{eq:surface-orders} and Appendix~\ref{app:surface} identifies \(1<\Gamma_{\rm env}<3/2\) as a sufficient range for the matter derivative correction to vanish. We construct a thermodynamically consistent \(C^2\) core-envelope completion, with continuous pressure and its first two derivatives in this range, with an exact enthalpy endpoint, and use it to define the vacuum boundary without a finite density cutoff.

The small EiBI corrections require accurate evaluation of both the interior equations at weak coupling and the exterior response at low compactness. We derive cancellation-free forms for these limits without changing the underlying equations. We compare EiBI and GR at fixed mass, then isolate the gravity residual for a chosen envelope. A separate envelope family is used to measure the variation of the absolute stellar predictions.

This paper is organized as follows. In Section~\ref{sec:gravity}, we introduce the EiBI field equations, metric conventions, and stellar background equations, while in Sec.~\ref{sec:tides} we discuss the physical surface quantity needed by the exterior matching. The core-envelope equation of state and its condition on $\Gamma_{\mathrm{env}}$ are developed in Sec.~\ref{sec:eos}, and the numerical formulation and its checks follow in Secs.~\ref{sec:numerics} and~\ref{sec:validation}. Sections~\ref{sec:results} and~\ref{sec:surface-systematics} present the fixed-mass comparisons and surface model sensitivity. In Section~\ref{sec:limitations}, we examine the surface prescription and observational scope, and Sec.~\ref{sec:conclusions} summarizes the principal findings. Appendix \ref{app:effective-fluid} to \ref{app:numerical-results}
provide some of the details of our computations. 

\section{EiBI Gravity and Conventions}
\label{sec:gravity}

We adopt the Palatini formalism of EiBI gravity with a torsion-free affine connection and the symmetric Ricci tensor \(R_{(\mu\nu)}(\Gamma)\). The Born-Infeld coupling \(\kappa\) has dimensions of area, and the constant \(\lambda\) sets the cosmological term. With matter fields \(\Psi\) and matter action \(S_{\rm m}\), the action in \(c=1\) units is
\begin{equation}
\begin{split}
 S={}&\frac{1}{8\pi G\kappa}\int d^4x
 \left[\sqrt{-\left|g_{\mu\nu}
 +\kappa R_{(\mu\nu)}(\Gamma)\right|}\right.\\
 &\left.-\lambda\sqrt{-g}\right]+S_{\rm m}[g,\Psi],
\end{split}
 \label{eq:eibi-action}
\end{equation}
Here, $|\cdots|$ denotes a determinant, with \(g=\det g_{\mu\nu}\) and \(q=\det q_{\mu\nu}\). The affine connection \(\Gamma^\alpha{}_{\mu\nu}\) is independent of \(g_{\mu\nu}\). Variation with respect to the independent fields gives an auxiliary metric \(q_{\mu\nu}\) compatible with \(\Gamma^\alpha{}_{\mu\nu}\) and the algebraic field equations
\cite{banados2010,delsate2012}
\begin{equation}
 q_{\mu\nu}=g_{\mu\nu}+\kappa R_{(\mu\nu)}(\Gamma),
 \qquad
 \frac{\sqrt{-q}}{\sqrt{-g}}q^{\mu\nu}
 =\lambda g^{\mu\nu}-8\pi G\kappa T^{\mu\nu}.
 \label{eq:eibi-field-equations}
\end{equation}
We set the cosmological constant to zero and therefore use \(\lambda=1\) throughout.

For a perfect fluid, \(T^\mu{}_\nu=\operatorname{diag}(-\epsilon,P,P,P)\), and minimal coupling to the physical metric gives \(\nabla^{(g)}_\mu T^{\mu\nu}=0\). The second relation in Eq.~\eqref{eq:eibi-field-equations} fixes the algebraic deformation between the two metrices, while the connection is the Levi-Civita connection of \(q_{\mu\nu}\). We evolve the differential equations in the auxiliary geometry and reconstruct observable radii and tidal data from the physical geometry. In vacuum, the matter deformation vanishes and 
GR is recovered.

To express these equations in stellar units, we retain SI lengths. For a pressure or energy density \(X\) in SI units, we define \(\widehat X\equiv GX/c^4\), which has units \(\mathrm{m}^{-2}\), so \(\kappa\widehat X\) is dimensionless. In particular, \(\Sigmahat\equiv G\Sigma/c^4\). We also define \(\mu(r)\equiv M(r)/\Msun\) and \(\RsunG\equiv G\Msun/c^2\). The interior \(M(r)\) is the auxiliary mass function, and its surface value gives the gravitational mass.

In the line elements below, \(x^0=ct\), and the coordinate \(r\) is the auxiliary radial coordinate. The dimensionless fluid factors \(a\) and \(b\), defined below, relate it to the physical circumferential radius. The line elements are 
\begin{align}
 ds_q^2=-e^{\beta}dt^2+e^{\alpha}dr^2+r^2d\Omega^2,
 \label{eq:auxiliary-metric}\\
 ds_g^2=-e^{\nu}dt^2+e^{\bar\lambda}dr^2+\frac{r^2}{ab}d\Omega^2.
 \label{eq:physical-metric}
\end{align}
The physical lapse, radial metric coefficient, and circumferential radius follow from
\begin{equation}
 e^\nu=e^\beta\frac{a}{b^3},\qquad
 e^{\bar\lambda}=\frac{e^\alpha}{ab},\qquad
 R_{\rm circ}=\frac{r}{\sqrt{ab}}.
 \label{eq:metric-map}
\end{equation}
At a regular vacuum surface, \(a,b\rightarrow 1\), so \(R_{\rm circ}=r\). Inside matter they generally differ at finite nonzero coupling.

The fluid factors depend on pressure and energy density. We define them together with the combinations used in the evolution equations:
\begin{align}
 a^2&=1+8\pi\kappa\epshat,
 \qquad b^2=1-8\pi\kappa\phat,
 \label{eq:ab-factors}\\
 A_1&=\frac{1}{ab}-1,
 \qquad A_2=\frac{a}{b^3}-1,\notag\\
 \mathcal A&=A_1+A_2,
 \label{eq:stable-combinations}\\
 e^{-\alpha}\equiv D&=1-\frac{2\RsunG\mu}{r},\notag\\
 \chi&=2\pi\kappa\left[
 \frac{3(\epshat+\phat)}{b^2}+\frac{\Sigmahat}{a^2}\right].
 \label{eq:chi-definition}
\end{align}
Here \(D\) is the Schwarzschild-like radial factor in the auxiliary radial coordinate. The quantities \(A_1\), \(A_2\) and \(\chi\) are all of order \(\kappa\) at weak coupling. 
\(A_1\) and \(A_2\) depend only on \(\epshat\) and \(\phat\), whereas \(\chi\) also contains the sound speed combination \(\Sigma\) from Eq.~\eqref{eq:sigma-definition}.
The relativistic enthalpy
\begin{equation}
 h(P)=\int_0^P\frac{d\widetilde P}{\epsilon(\widetilde P)+\widetilde P}
 \label{eq:enthalpy-definition}
\end{equation}
is monotone along the interior solution of each stellar model. We define \(r_s\) by the first downward root \(h(r_s)=0\), the gravitational mass as \(M\equiv\Msun\mu(r_s)\), and the physical radius as \(R\equiv R_{\rm circ}(r_s)\). Thus the thermodynamic coordinate \(h\) locates the vacuum boundary directly. For every finite nonzero \(\kappa\), the background equations are
\begin{align}
 \frac{d\mu}{dr}
 &=\frac{r^2}{4\RsunG\kappa}(A_2-3A_1),
 \label{eq:eibi-mass}\\
 \frac{dh}{dr}
 &=-\frac{\RsunG\mu/r^2+r\mathcal A/(4\kappa)}{(1+\chi)D}.
 \label{eq:eibi-enthalpy}
\end{align}
The weak-coupling limit \(\kappa\rightarrow0\) is regular:
\begin{align}
 A_2-3A_1&=16\pi\kappa\epshat+O(\kappa^2),\notag\\
 \mathcal A&=16\pi\kappa\phat+O(\kappa^2),
 \qquad \chi=O(\kappa).
 \label{eq:background-gr-limit}
\end{align}
Although these limits are finite, direct subtraction in their numerators can lose the small terms that survive division by \(\kappa\). Section~\ref{sec:numerics} gives stable finite-coupling expressions. At exactly \(\kappa=0\), we instead use the GR equations:
\begin{align}
 \frac{d\mu}{dr}&=\frac{4\pi r^2\epshat}{\RsunG},
 \label{eq:gr-mass}\\
 \frac{dh}{dr}&=-\frac{\RsunG\mu/r^2+4\pi r\phat}{D}.
 \label{eq:gr-enthalpy}
\end{align}
Vacuum is a separate explicit branch with \(d\mu/dr=0\). A regular stellar solution requires \(a^2>0\), \(b^2>0\), \(D>0\), and \(1+\chi>0\) before the surface is reached. Violation of any condition terminates the calculation. Subject to these conditions, the background solution determines \(M\) and \(R\).

\section{Quadrupolar Perturbations and Physical Matching}
\label{sec:tides}

The equilibrium solution provides the background for a linear, static, even-parity tide. In Regge-Wheeler gauge, the scalar spherical harmonic \(Y_{\ell m}\) carries the angular dependence. Spherical symmetry makes the radial equations independent of \(m\). A prime denotes \(d/dr\). We denote the auxiliary and physical temporal amplitudes by \(F\) and \(H\), respectively, and their radial and angular counterparts by \(L_{q,g}\) and \(K_{q,g}\). The perturbed line elements are given by the expressions \cite{hinderer2008}
\begin{align}
 d\widetilde s_q^2={}&-e^\beta(1+F Y_{\ell m})dt^2\notag\\
 &+e^\alpha(1+L_qY_{\ell m})dr^2\notag\\
 &+r^2(1+K_qY_{\ell m})d\Omega^2,
 \label{eq:perturbed-q}\\
 d\widetilde s_g^2={}&-e^\nu(1+H Y_{\ell m})dt^2\notag\\
 &+e^{\bar\lambda}(1+L_gY_{\ell m})dr^2\notag\\
 &+\frac{r^2}{ab}(1+K_gY_{\ell m})d\Omega^2.
 \label{eq:perturbed-g}
\end{align}
Eliminating the other metric amplitudes leaves a second-order equation for the auxiliary temporal amplitude \(F\). Its overall normalization is arbitrary, so we instead evolve the logarithmic derivative \(y_F\). This converts the second-order equation into a first-order nonlinear equation (the Riccati form),
\begin{align}
 F''+fF'+gF&=0,
 \qquad y_F\equiv\frac{rF'}{F},\notag\\
 y_F'&=\frac{y_F-y_F^2}{r}-fy_F-rg.
 \label{eq:riccati}
\end{align}
Regularity at the center requires \(F(r)=F_0r^\ell[1+O(r^2)]\), and therefore \(y_F(r)=\ell+O(r^2)\). We set \(\ell=2\) for the quadrupole response.

To write the coefficients in terms of the background solution at finite nonzero \(\kappa\), we use the auxiliary lapse derivative and matter combination
\begin{equation}
 \beta'=\frac{r\mathcal A/(2\kappa)+2\RsunG\mu/r^2}{D},
 \qquad
 \mathcal Q=A_2+\frac{a}{b^3}\frac{\chi}{1+\chi}.
 \label{eq:tidal-intermediates}
\end{equation}
The finite-coupling perturbation coefficients are then
\begin{align}
 f&=\frac{rA_1/\kappa+1/r}{D}+\frac{1}{r},
 \label{eq:eibi-f}\\
 g&=\frac{2\mathcal Q/\kappa-\ell(\ell+1)/r^2}{D}-(\beta')^2.
 \label{eq:eibi-g}
\end{align}
The term \(-(\beta')^2\) arises from the radial variation of the auxiliary lapse and remains outside the factor \(D^{-1}\) that multiplies the matter and angular terms. The weak-coupling limits are
\begin{equation}
 \frac{A_1}{\kappa}\longrightarrow4\pi(\phat-\epshat),
 \qquad
 \frac{\mathcal Q}{\kappa}\longrightarrow
 2\pi(5\epshat+9\phat+\Sigmahat)
 \label{eq:tidal-gr-limit}
\end{equation}
Together, these limits recover the exact GR coefficients
\begin{align}
 f_{\rm GR}&=\frac{2}{r}
 +\frac{2\RsunG\mu/r^2+4\pi r(\phat-\epshat)}{D},
 \label{eq:gr-f}\\
 \beta'_{\rm GR}&=
 \frac{2(\RsunG\mu/r^2+4\pi r\phat)}{D},
 \label{eq:gr-beta}\\
 g_{\rm GR}&=\frac{-\ell(\ell+1)/r^2
 +4\pi(5\epshat+9\phat+\Sigmahat)}{D}\notag\\
 &\quad-(\beta'_{\rm GR})^2.
 \label{eq:gr-g}
\end{align}
These are the established static tidal equations and their EiBI mapping \cite{hinderer2008,hinderer2009erratum,sham2014,prasetyo2021}.

Exterior matching must be distinguished from regularity of the physical metric. For a surface without a material shell, the EiBI junction conditions impose continuity of the induced physical and auxiliary metrics and of the auxiliary extrinsic curvature, \([g_{ij}]=[q_{ij}]=[K_{ij}(q)]=0\), as given in Eq.~(74) of Ref.~\cite{pani2012phenomenology}. They do not impose the GR extrinsic curvature condition on \(g_{\mu\nu}\) inside matter. Since \(q_{\mu\nu}=g_{\mu\nu}\) in vacuum, the auxiliary tidal solution can give a value for matching to the Schwarzschild exterior. A regular auxiliary solution alone does not establish regularity of the physical metric to which matter couples \cite{pani2012surface}. We therefore also examine the physical perturbation and its derivative as the surface is approached from inside. Linearizing the algebraic metric relation and eliminating the barotropic fluid perturbation, as derived in Appendix~\ref{app:effective-fluid}, gives the amplitude map \cite{prasetyo2021} and its logarithmic derivative with respect to auxiliary radius:
\begin{equation}
 \frac{H}{F}=\frac{1}{1+\chi},\qquad
 y_H^{(r)}=y_F-\frac{r\chi'}{1+\chi}.
 \label{eq:amplitude-map}
\end{equation}
An amplitude ratio approaching unity does not ensure equality of its derivatives: a residual gradient can survive at the surface. A second conversion is also needed because Schwarzschild matching uses the physical circumferential radius. From Eq.~\eqref{eq:metric-map},
\begin{equation}
 \frac{dR_{\rm circ}}{dr}=\frac{1}{\sqrt{ab}}
 \left[1-\frac{r}{2}\left(\frac{a'}a+\frac{b'}b\right)\right].
 \label{eq:physical-radius-derivative}
\end{equation}
After applying the amplitude map, the radial coordinate chain rule gives
\begin{equation}
 y_H^{(R)}\equiv\frac{R_{\rm circ}}{H}\frac{dH}{dR_{\rm circ}}
 =\frac{R_{\rm circ}/r}{dR_{\rm circ}/dr}
 \left(y_F-\frac{r\chi'}{1+\chi}\right).
 \label{eq:physical-y-map}
\end{equation}

The low-density EOS determines the vacuum limit of Eq.~\eqref{eq:physical-y-map}. For the regulated surface model of Sec.~\ref{sec:eos}, pressure obeys \(P\sim\rho^{\Gamma_{\rm env}}\), where \(\rho\) is rest-mass density. Under the surface assumptions derived in Appendix~\ref{app:surface},
\begin{equation}
 \chi=O(\rho^{2-\Gamma_{\rm env}}),\qquad
 r\chi'=O(\rho^{3-2\Gamma_{\rm env}}).
 \label{eq:surface-orders}
\end{equation}
In the same limit, the radius map gives \(R_{\rm circ}/r\rightarrow1\) and \(dR_{\rm circ}/dr\rightarrow1\). It follows that
\begin{equation}
\begin{split}
 1<\Gamma_{\rm env}<\frac{3}{2}
 \quad\Longrightarrow\quad&
 \lim_{r\to r_s^-}y_H^{(R)}[R_{\rm circ}(r)]\\
 &=\lim_{r\to r_s^-}y_F(r)\equiv y_s.
\end{split}
 \label{eq:surface-equality}
\end{equation}
This range is sufficient under the assumptions of Appendix~\ref{app:surface}. At \(\Gamma_{\rm env}=3/2\), \(r\chi'\) tends to a finite EOS-dependent constant, so a finite correction remains. For \(3/2<\Gamma_{\rm env}<2\) the correction diverges, and stiffer surface laws need a separate analysis. This dependence on matter derivatives is the surface sensitivity known in Palatini theories \cite{pani2012surface}.

The origin of this correction is not specific to EiBI gravity. In Ricci-based theories formulated in the Palatini formalism, the physical metric is an algebraic function of the auxiliary metric and the local stress-energy \cite{afonso2018}. Palatini \(f(\mathcal R)\) gravity gives a concrete example. There the trace of the field equations, \(\mathcal R f_{\mathcal R}-2f=8\pi\widehat T\) with \(\widehat T=3\phat-\epshat\), fixes \(\mathcal R\) algebraically in terms of the matter, and the independent connection is the Levi-Civita connection of \(h_{\mu\nu}=f_{\mathcal R}\,g_{\mu\nu}\) \cite{olmo2011}. Repeating the steps of Appendix~\ref{app:effective-fluid} with \(g_{\mu\nu}=h_{\mu\nu}/f_{\mathcal R}\), \(\delta\widehat T=\delta\phat\,(3-1/c_s^2)\) and Eq.~\eqref{eq:static-fluid-amplitude} gives \(H/F=1/(1+\chi_f)\), with
\begin{equation}
 \chi_f=\frac{4\pi f_{\mathcal R\mathcal R}\,[\Sigmahat-3(\epshat+\phat)]}
 {f_{\mathcal R}\,(\mathcal R f_{\mathcal R\mathcal R}-f_{\mathcal R})}.
 \label{eq:palatini-chi}
\end{equation}
For \(f=\mathcal R+\alpha\mathcal R^2\), the trace equation gives \(\mathcal R=-8\pi\widehat T\) and \(\chi_f=-8\pi\alpha[\Sigmahat-3(\epshat+\phat)]/(1+2\alpha\mathcal R)\). As for \(\chi\) in Eq.~\eqref{eq:chi-definition}, the term in \(\Sigmahat\) dominates near the surface. For models with \(f_{\mathcal R\mathcal R}(0)\neq0\), such as \(f=\mathcal R+\alpha\mathcal R^2\), \(\chi_f\) then has the orders of Eq.~\eqref{eq:surface-orders}, while \(f_{\mathcal R}\to1\) and the circumferential radius \(r/\sqrt{f_{\mathcal R}}\) tends to \(r\). The condition in Eq.~\eqref{eq:surface-equality} therefore carries over to these models. If \(f_{\mathcal R\mathcal R}(0)=0\), \(\chi_f\) vanishes faster at the surface and the threshold on \(\Gamma_{\rm env}\) moves to larger values. For polytropic surfaces, the background curvature in that theory which satisfy the relevant nondegeneracy conditions, also diverges for \(3/2<\Gamma<2\) \cite{barausse2008nogo,barausse2008curvature}. A full tidal calculation in Palatini \(f(\mathcal R)\) gravity is beyond the scope of this paper.

In the code, we  check that $\Gamma_{\rm env}$ lies in this range and then use the limit of Eq.~\eqref{eq:surface-equality}. For \(\kappa=0\), \(\chi\) is exactly zero, and the auxiliary and physical variables coincide without the finite-coupling surface argument.

For our regular envelope, Eq.~\eqref{eq:surface-equality} identifies their common boundary derivative, so Schwarzschild matching uses \((C,y_s)\) \cite{sterne1939,hinderer2008,damour2009}. At the surface defined by \(h(r_s)=0\), we use the physical compactness \(C=\RsunG\mu(r_s)/R\) and set \(y=y_s\):
\begin{equation}
 k_2=\frac{8}{5}C^5(1-2C)^2
 \frac{2+2C(y-1)-y}{\mathcal D_k},
 \label{eq:love-exact}
\end{equation}
The denominator is
\begin{align}
 \mathcal D_k={}&2C[6-3y+3C(5y-8)]\notag\\
 &+4C^3[13-11y+C(3y-2)+2C^2(1+y)]\notag\\
 &+3(1-2C)^2[2+2C(y-1)-y]\ln(1-2C).
 \label{eq:love-denominator}
\end{align}
The dimensionless deformability follows from Eq.~\eqref{eq:tidal_def}.

We therefore need an EOS that lets the star reach zero density with a finite, well-defined \(y_s\), without a density cutoff.

\section{Core and Regulated Envelope Equation of State}
\label{sec:eos}

We retain the zero-temperature Chandrasekhar electron gas in the cold core \cite{chandrasekhar1931}. In this idealization, ions provide most of the rest-mass density, and degenerate electrons provide the pressure. We omit ionic thermal pressure, Coulomb corrections, and composition-dependent reaction thresholds. The neutral matter mass per electron is \(\mu_e m_u\), with \(\mu_e=2\). This value is approximately appropriate for fully ionized helium, carbon, or oxygen matter without fixing a detailed composition. Here \(m_u\) is the atomic mass unit. To express the EOS, define \(x=p_F/(m_ec)\), \(s=\sqrt{1+x^2}\), and \(n_e=p_F^3/(3\pi^2\hbar^3)\), where \(p_F\), \(m_e\), and \(n_e\) are the electron Fermi momentum, mass, and number density, and \(\hbar\) is the reduced Planck constant. The density and pressure scales are
\begin{align}
 \rho&=\mu_e m_un_e=\rho_\star x^3,\notag\\
 \rho_\star&=\mu_e m_u\frac{(m_ec)^3}{3\pi^2\hbar^3},
 \qquad
 P_\star=\frac{m_e^4c^5}{24\pi^2\hbar^3}.
 \label{eq:eos-scales}
\end{align}
The pressure and total energy density are
\begin{align}
 P_{\rm core}(x)&=P_\star
 [x(2x^2-3)s+3\operatorname{asinh}x],
 \label{eq:core-pressure}\\
 \epsilon_{\rm core}(x)&=\rho c^2+3P_\star
 \left[x(1+2x^2)s-\operatorname{asinh}x-\frac{8}{3}x^3\right].
 \label{eq:core-energy}
\end{align}
The subtraction in the energy density isolates the electron kinetic energy. Because the neutral matter rest mass already contains the electron rest energy, adding it again would double count that contribution.

At low density, where \(x\ll1\), the core EOS has the expansions
\begin{align}
 P_{\rm core}&=\frac{8}{5}P_\star x^5+O(x^7)
 \propto\rho^{5/3},\notag\\
 \epsilon_{\rm core}&=\rho c^2+\frac{3}{2}P_{\rm core}
 +O(\rho^{7/3}).
 \label{eq:core-low-density}
\end{align}
The exponent \(5/3\) lies in the range where the finite-coupling derivative correction in Eq.~\eqref{eq:surface-orders} diverges under the assumptions of Appendix~\ref{app:surface}. Matching the auxiliary tidal solution to the Schwarzschild exterior does not remove this physical derivative singularity. We replace the low-density edge to obtain the regular physical surface required by our model.

The replacement joins a polytropic envelope to the core over one decade in density. Let \(\rho_{\rm tr}\) be the upper transition density and \(\rho_-=\rho_{\rm tr}/10\) the lower one. The analytic envelope pressure is
\begin{equation}
 P_{\rm env}=K\rho^{\Gamma_{\rm env}},\qquad
 K=\frac{P_{\rm core}(\rho_{\rm tr})}
 {\rho_{\rm tr}^{\Gamma_{\rm env}}},
 \label{eq:envelope-pressure}
\end{equation}
The composite pressure equals \(P_{\rm env}\) for \(\rho\leq\rho_-\) and \(P_{\rm core}\) for \(\rho\geq\rho_{\rm tr}\). Between these joins, we blend the logarithm of the pressure across the same density decade. With \(u_\rho=\ln\rho\), evaluated in a fixed density unit, and the dimensionless blend coordinate \(t_\rho\),
\begin{align}
 \ln P&=(1-S)\ln P_{\rm env}+S\ln P_{\rm core},
 \label{eq:pressure-blend}\\
 S(t_\rho)&=6t_\rho^5-15t_\rho^4+10t_\rho^3,\qquad
 t_\rho=\frac{u_\rho-\ln\rho_-}{\ln\rho_{\rm tr}-\ln\rho_-}.
 \label{eq:smoothstep}
\end{align}
Blending logarithmic pressure preserves positivity. The quintic function has vanishing first and second derivatives at both endpoints, making \(P(\rho)\) twice continuously differentiable across the joins. This matters for the tidal map: thermodynamic consistency gives
\begin{equation}
 c_s^2=\frac{\rho P_{,\rho}}{\epsilon+P},\qquad
 \Sigma=\frac{(\epsilon+P)^2}{\rho P_{,\rho}},
 \label{eq:sigma-pressure-derivatives}
\end{equation}
Here \(,\rho\) denotes differentiation with respect to density. The differentiated tidal map depends on \(P_{,\rho\rho}\) as well as \(P_{,\rho}\). The \(C^2\) blend avoids an artificial jump in that theoretical map at either join. 

We obtain the energy density \(\epsilon\) from the blended pressure, rather than blending the two independently. For this cold barotropic fluid, pressure and energy density depend only on \(\rho\) and must satisfy the first law of thermodynamics,
\begin{equation}
 \frac{d\epsilon}{d\rho}=\frac{\epsilon+P}{\rho},
 \qquad
 \frac{d}{d\ln\rho}\left(\frac{\epsilon}{\rho}\right)=\frac{P}{\rho},
 \label{eq:thermodynamic-reconstruction}
\end{equation}
The matched core value \(\epsilon_{\rm core}(\rho_{\rm tr})\) provides the
upper boundary condition. We then integrate the first law relation downward in density through the blend:
\begin{equation}
 \frac{\epsilon(\rho)}{\rho}
 =\frac{\epsilon_{\rm core}(\rho_{\rm tr})}{\rho_{\rm tr}}
 +\int_{\ln\rho_{\rm tr}}^{\ln\rho}
 \frac{P(e^{u_\rho})}{e^{u_\rho}}\,du_\rho .
 \label{eq:energy-density-integral}
\end{equation}
The analytic surface branch has
\begin{equation}
 \epsilon=e_0\rho+\frac{P}{\Gamma_{\rm env}-1},\qquad
 h=\ln\left[1+\frac{\Gamma_{\rm env}K
 \rho^{\Gamma_{\rm env}-1}}
 {(\Gamma_{\rm env}-1)e_0}\right].
 \label{eq:surface-eos}
\end{equation}
Continuity of the energy density at \(\rho_-\) fixes the remaining coefficient:
\begin{equation}
 e_0=\frac{\epsilon(\rho_-)
 -P(\rho_-)/(\Gamma_{\rm env}-1)}
 {\rho_-},
 \label{eq:surface-e0}
\end{equation}
Here \(\epsilon(\rho_-)\) is the value reached by the downward integration of Eq.~\eqref{eq:thermodynamic-reconstruction}. Thus \(e_0\) is fixed by the core matching, rather than chosen independently as an envelope normalization.

The surface relation in Eq.~\eqref{eq:surface-eos} also gives the exact analytic inverse
\begin{equation}
 \rho(h)=\left[
 \frac{(\Gamma_{\rm env}-1)e_0}{\Gamma_{\rm env}K}
 \left(e^h-1\right)
 \right]^{1/(\Gamma_{\rm env}-1)}.
 \label{eq:surface-enthalpy-inverse}
\end{equation}
We evaluate \(e^h-1\) in a form that stays accurate as \(h\to0\), so the analytic inverse gives exactly zero density at \(h=0\). In the blend and core, the baseline calculation instead uses a monotone piecewise cubic interpolation of \(u_\rho=\ln\rho\) as a function of \(h\), with 256 points per density decade \cite{fritsch1984}. 

This construction gives a regular surface for matching, but it does not select a unique atmosphere. To measure its model dependence, we vary the surface index and transition density, combining \(\Gamma_{\rm env}\in\{1.25,4/3,1.45\}\) with \(\rho_{\rm tr}\in\{10^2,10^3,10^4\}\,\mathrm{g\,cm^{-3}}\). We solve the stellar equations separately for each of these nine envelopes.

The baseline envelope is \(\Gamma_{\rm env}=4/3\) and \(\rho_{\rm tr}=10^3\,\mathrm{g\,cm^{-3}}\). The spread across these models measures sensitivity to the envelope choice.
The blend width and quintic function are held fixed. 

\section{Cancellation-Free Numerics}
\label{sec:numerics}

This section describes the baseline calculation in double precision. Appendix~\ref{app:hp} gives the separate high-precision method used for the much smaller variation of the EiBI-GR difference between envelopes. That method refines the EOS and stellar integration and evaluates the closed exterior formula at its working precision.

Subtracting nearly equal rounded numbers can erase a small difference. This occurs in the degenerate gas formulas at small \(x\), the EiBI metric factors near \(\kappa=0\), and the exterior Love formula at low compactness. For \(x\leq0.1\), we evaluate the EOS formulas from their Taylor expansions. For \(0.1<x\leq6\), we use series with positive coefficients in \(\tau=\tanh[\operatorname{asinh}(x)/2]\), and the closed forms are used at larger \(x\). For finite nonzero \(\kappa\), we evaluate \(\ln(1+x)\) and \(e^x-1\) in forms that stay accurate for small \(x\), together with the ratios in Appendix~\ref{app:stable-algebra}. At exactly \(\kappa=0\), we use the GR equations without dividing by the coupling.

In the exterior formula, the numerator and denominator both vanish as high powers of compactness, although their ratio has a finite limit. Direct evaluation of Eqs.~\eqref{eq:love-exact} and \eqref{eq:love-denominator} therefore loses precision for white dwarfs. For \(C\leq 0.045\), we remove the common small factor analytically and use the rearrangement truncated at tenth order,
\begin{equation}
 k_2=-\frac{3}{2}
 \frac{(y-2)+2C(1-y)}
 {\mathcal U_{10}(C)+y\mathcal V_{10}(C)},
 \label{eq:love-low-c}
\end{equation}
where \(\mathcal U_{10}(C)=\sum_{n=0}^{10}u_nC^n\) and \(\mathcal V_{10}(C)=\sum_{n=0}^{10}v_nC^n\). The tenth-order coefficient tuples, listed in ascending powers of \(C\), are
\begin{align}
 (v_n)_{n=0}^{10}={}&
 (3,9,150/7,330/7,100,208,\notag\\
 &\hspace{1.4cm}4704/11,9600/11,23040/13,\notag\\
 &\hspace{1.4cm}326400/91,50688/7),
 \label{eq:love-b-coefficients}\\
 (u_n)_{n=0}^{10}={}&
 (9,36,750/7,1980/7,700,1664,\notag\\
 &\hspace{1.4cm}42336/11,96000/11,253440/13,\notag\\
 &\hspace{1.4cm}3916800/91,658944/7).
 \label{eq:love-a-coefficients}
\end{align}
The coefficients satisfy \(u_n=(n+3)v_n\). The rearranged form also gives the Newtonian limit,
\begin{equation}
 \lim_{C\to0}k_2=\frac{2-y}{2(3+y)},
 \label{eq:love-newtonian-limit}
\end{equation}
which is consistent with the Clairaut-Radau convention used in the Lane-Emden comparison below. For \(C>0.045\), we evaluate the complete exterior formula. We checked that the two forms agree at \(C=0.045\) for \(0.75\leq y\leq2.0\). Every white dwarf model in the specified grid lies well below this switch. 

The baseline calculation evolves \((h,\mu,y_F)\) with the auxiliary radius \(r\). Because the coordinate equations are singular at \(r=0\), we use the center expansion in Appendix~\ref{app:center} to initialize the integration at \(r_0=5\times10^{-4}\,\mathrm m\). We use SciPy's DOP853 method \cite{scipy2020}, with relative tolerance \(10^{-13}\), absolute tolerance \(10^{-15}\) for each of the three variables, maximum step \(2500\,\mathrm m\), and maximum radius \(3\times10^8\,\mathrm m\). The integration stops where \(h\) reaches zero, not at a finite pressure. If a trial step overshoots to \(h<0\), the matter terms are set to their vacuum values.

At the surface we require \(|h(r_s)|\leq10^{-15}\). We also compare its state with the interpolation within the accepted step (dense output), requiring
\begin{equation}
\begin{split}
 |z_i^{\rm dense}-z_i^{\rm event}|
 \leq 8\bigl[{\rm atol}_i+{\rm rtol}\,
 \max\bigl(&|z_i^{\rm dense}|,\\
 &|z_i^{\rm event}|\bigr)\bigr],
\end{split}
 \label{eq:event-dense-consistency}
\end{equation}
for \(z_i\in\{h,\mu,y_F\}\). We check that the located surface agrees with the solver's interpolated solution. There \(a^2=b^2=1\) and \(\chi=0\), so \(y_H^{(R)}(R)=y_F(r_s)\), and we compute \(C\), \(k_2\), and \(\Lambda\).

To compare stars at fixed mass, we adjust the central density \(\rho_c\) until the gravitational mass equals the target \(M_\star\). We find \(\rho_c\) with a secant method safeguarded by bisection, starting from a bracket on the coarse grid, and stop when \(|M-M_\star|\leq10^{-12}\,\Msun\) and \(\Delta\log_{10}\rho_c\le0.02\) both are satisfied.

We use five illustrative couplings
\begin{equation}
 \kappa\in\{-5\times10^6,-10^6,0,10^6,5\times10^6\}\,\mathrm{m^2},
 \label{eq:kappa-grid}
\end{equation}
the nine envelope models described above, and a 21-point logarithmic coarse central density grid. We refine sequence extrema from brackets formed by computed sequence points.
Each fixed-mass model is a separate stellar computation and so no interpolation is involved. The complete model grid is reported in Table~\ref{tab:model-grid}.

\begingroup
\squeezetable
\begin{table*}
\caption{Sampled model parameters. Here \(\kappa\) is the gravity coupling, \(\Gamma_{\rm env}\) the envelope polytropic exponent, \(\rho_{\rm tr}\) its transition density, and \(\rho_c\) the central rest-mass density. All combinations of the three exponents and three transition densities give nine envelopes. The 21 logarithmically spaced densities form the initial sequence grid for each coupling and envelope. The mass maxima are refined separately. Count denotes the number of sampled values on each axis.}
\label{tab:model-grid}
\begin{ruledtabular}
\begin{tabular}{lrc}
Axis & Values & Count \\
\hline

$\kappa\,[\mathrm{m}^2]$ & $-5\times 10^{6}, -1\times 10^{6}, 0, 1\times 10^{6}, 5\times 10^{6}$ & 5 \\

$\Gamma_{\rm env}$ & $1.25, 4/3, 1.45$ & 3 \\

$\rho_{\rm tr}\,[\mathrm{g\,cm^{-3}}]$ & $100, 1000, 10000$ & 3 \\

$\rho_c\,[\mathrm{g\,cm^{-3}}]$ & $1\times 10^{6}$--$1\times 10^{11}$ (log-spaced) & 21 \\

$M/M_\odot$ & $0.6, 0.8, 1, 1.2, 1.3$ & 5 \\
\end{tabular}
\end{ruledtabular}
\end{table*}
\endgroup

\section{Validation and Numerical Accuracy}
\label{sec:validation}

We test the calculation in three ways: against known limits, against an independent GR code, and by refining the numerical resolution. The Newtonian \(n=3/2\) Lane-Emden polytrope tests the tidal normalization for a nonrelativistic degenerate electron gas. Its calculated apsidal constant is \(k_2=0.14327877\), consistent with the tabulated value of \cite{brooker1955}. If \(\eta_2(R)\) is the surface solution of the Clairaut-Radau equation, the convention is
\begin{equation}
 k_2^{\rm N}=\frac{3-\eta_2(R)}{2[2+\eta_2(R)]},
 \label{eq:clairaut-radau-k2}
\end{equation}
consistent with Eq.~\eqref{eq:love-newtonian-limit} through \(\eta_2(R)=y_s+1\) for a surface with vanishing density.  

To test the relativistic equations and physical matching, the GR comparison uses \(\rho_c=10^9\,\mathrm{g\,cm^{-3}}\), \(\kappa=0\), and the baseline envelope. Both calculations use the same regulated EOS map. The reference implementation independently constructs the regular center data, evolves the GR background and tidal equations, locates the zero-enthalpy surface, computes the physical radius, and applies a high-precision Schwarzschild matching. Agreement with the DOP853 \(\kappa=0\) branch tests the equation and exterior matching.

A separate comparison changes only the integration method from DOP853 to Radau. The regulated EOS, regular center state, background and tidal equations, zero-enthalpy event, and exterior matching are shared. It therefore tests integration and event location sensitivity. The Radau settings for both checks are given in Appendix~\ref{app:validation}.

We assessed finite-coupling resolution for the model with \(\rho_c=10^9\,\mathrm{g\,cm^{-3}}\), \(\kappa=+10^6\,\mathrm{m^2}\), \(\Gamma_{\rm env}=4/3\), and \(\rho_{\rm tr}=10^3\,\mathrm{g\,cm^{-3}}\). Relative to the baseline solve, the refined solve divides each local error tolerance by 4.5 both the maximum step and center radius (to \(1250\,\mathrm m\) and \(2.5\times10^{-4}\,\mathrm m\)). 

We compare two calculations \(a\) and \(b\) of a quantity \(X\) through the symmetric relative difference \(s(X)=2|X_a-X_b|/(|X_a|+|X_b|)\), which we use throughout. At this model, \(s\) is \(3.08\times 10^{-13}\) in mass, \(4.63\times 10^{-13}\) in radius, \(8.46\times 10^{-12}\) in \(k_2\), and \(1.23\times 10^{-11}\) in \(\Lambda\) (Table~\ref{tab:convergence}). All are well below the tolerances listed there.

We repeated every fixed-mass solution with a quarter of each tolerance and half the maximum step and starting radius. The residuals \(\delta_X\) and half ranges \(\mathcal H_X\) defined in Secs.~\ref{sec:results} and~\ref{sec:surface-systematics} change by at most \(0.00154\) and \(0.000901\) respectively, well below \(1\%\).
Table~\ref{tab:gr-benchmarks} reports selected analytical and independent GR checks.
The comparison with the Perot-Chamel white dwarf fit is EOS-dependent because the computation includes microphysics absent from the present idealized EOS \cite{perot2022}. The differences mainly reflect the different EOS and are shown for reference. 

\begingroup
\squeezetable
\begin{table}[tb]
\caption{Numerical refinement of one EiBI model: \(\rho_c=10^9\,\mathrm{g\,cm^{-3}}\), \(\kappa=10^6\,\mathrm{m^2}\), \(\Gamma_{\rm env}=4/3\), and \(\rho_{\rm tr}=10^3\,\mathrm{g\,cm^{-3}}\). \(s(X)\) is defined in Sec.~\ref{sec:validation}. Solve \(b\) uses fourfold tighter error tolerances and half the maximum step and starting radius of solve \(a\).}
\label{tab:convergence}
\begin{ruledtabular}
\begin{tabular}{lrr}
Observable & $s(X)$ & Tolerance \\
\hline
$M$ & $3.08\times 10^{-13}$ & $1\times 10^{-7}$ \\
$R$ & $4.63\times 10^{-13}$ & $1\times 10^{-6}$ \\
$k_2$ & $8.46\times 10^{-12}$ & $1\times 10^{-5}$ \\
$\Lambda$ & $1.23\times 10^{-11}$ & $5\times 10^{-5}$ \\
\end{tabular}
\end{ruledtabular}
\end{table}
\endgroup

\section{Equilibrium Sequences and Fixed-Mass EiBI Response}
\label{sec:results}

\begin{figure}[!tbp]
\centering
\includegraphics[width=\columnwidth]{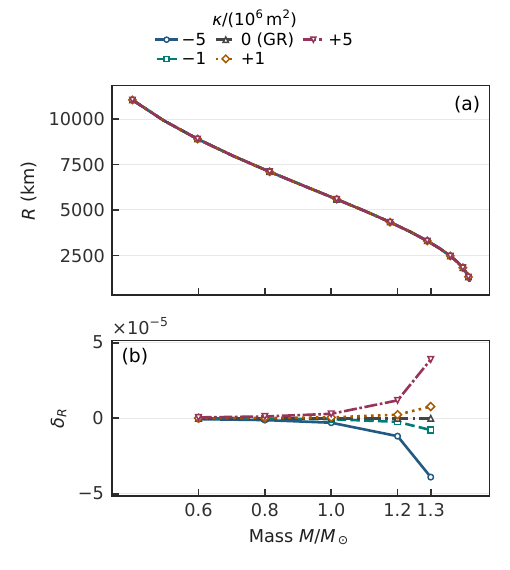}
\caption{(a): mass-radius sequences for the baseline envelope and sampled couplings, restricted to contiguous pre-turning-point segments that reach the vacuum surface. Radius decreases as central density and mass increase. 
(b): fixed-mass fractional residual \(\delta_R\) from Eq.~\eqref{eq:fractional-residual} at the five directly calculated mass targets.}
\label{fig:mass-radius}
\end{figure}

\begin{figure}[!tbp]
\centering
\includegraphics[width=\columnwidth]{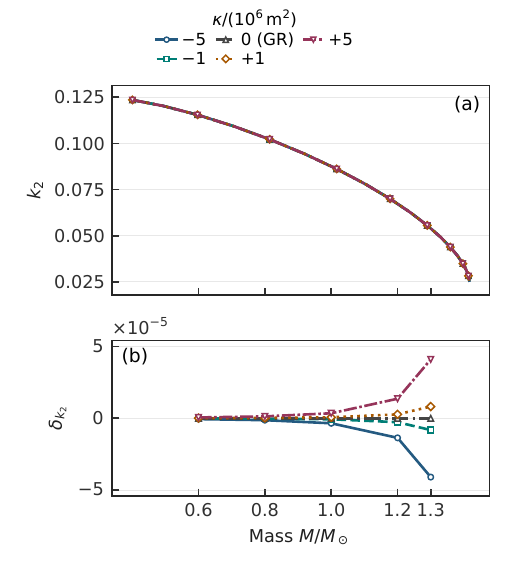}
\caption{(a): quadrupole Love number vs mass along the pre-turning-point sequence segments for the baseline envelope and sampled couplings. The absolute response decreases with mass. 
(b): fixed-mass residual \(\delta_{k_2}\) at the five calculated targets.}
\label{fig:k2-mass}
\end{figure}

\begin{figure}[!tbp]
\centering
\includegraphics[width=\columnwidth]{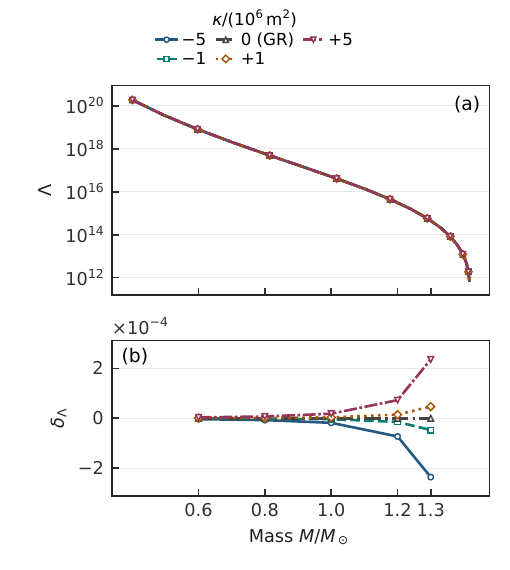}
\caption{(a): dimensionless tidal deformability vs mass along the same pre-turning-point sequence segments as Fig.~\ref{fig:k2-mass}. The logarithmic scale shows the strong decrease associated with increasing compactness. 
(b): \(\delta_\Lambda\) at the five calculated masses. The coupling ordering is unresolved in the absolute panel.}
\label{fig:lambda-mass}
\end{figure}

We first hold the envelope at its baseline choice and vary gravity. The upper panels of Figs.~\ref{fig:mass-radius}-\ref{fig:lambda-mass} show the pre-turning-point equilibrium sequences as central density increases. The coupling differences are too small to resolve on these absolute scales. The lower panels isolate them at five fixed masses between \(0.6\) and \(1.3\,\Msun\). Each target is a direct central density solve ending at \(h=0\). We define the fractional difference from GR by
\begin{equation}
 \delta_X(M,\kappa)\equiv
 \frac{X(M,\kappa)-X(M,0)}{X(M,0)}
 \label{eq:fractional-residual}
\end{equation}
with \(X\in\{R,k_2,\Lambda\}\) and the same baseline envelope in both models. We use the argument order \((M,\kappa)\) throughout. Results are computed only at the five target masses. Envelope variation is assessed separately in Fig.~\ref{fig:surface-systematics}.

Over the sampled baseline envelope grid, the largest absolute residuals are \(3.89\times 10^{-5}\) in radius, \(4.09\times 10^{-5}\) in Love number, and \(2.36\times 10^{-4}\) in deformability.

At fixed gravitational mass, the three residuals obey an exact identity. Since
\(C(\kappa)/C_{\rm GR}=R_{\rm GR}/R(\kappa)\),
\begin{equation}
 1+\delta_\Lambda
 =(1+\delta_{k_2})(1+\delta_R)^5.
 \label{eq:residual-identity}
\end{equation}
For small deviations from GR,
\begin{equation}
 \delta_\Lambda=\delta_{k_2}+5\delta_R+O(\delta^2).
 \label{eq:residual-linearized}
\end{equation}
The deformability residual is therefore not independent of the other two: the radius change enters with a factor of five to leading order.

Figure~\ref{fig:mass-radius} shows radius decreasing as mass increases along each displayed sequence, with a steeper decline near the upper end. This is the familiar compression of a degenerate white dwarf: as electron momenta increase, the core pressure in Eq.~\eqref{eq:core-pressure} approaches the relativistic degeneracy law and grows less rapidly with density. The approach to a limiting mass for an ideal white dwarf follows from relativistic electron degeneracy \cite{chandrasekhar1931}. The additional result here is the small, resolved displacement at equal gravitational mass. At every sampled target, negative \(\kappa\) contracts the baseline model relative to GR and positive \(\kappa\) expands it. The magnitudes increase toward the high-mass end.

Figure~\ref{fig:k2-mass} removes the explicit fifth-power radius scale from the tidal response. The absolute Love number decreases throughout each displayed segment. This trend is encoded in the increasing surface logarithmic derivative: Eq.~\eqref{eq:love-newtonian-limit} makes the leading low-compactness response a decreasing function of \(y_s\). It carries information about the density and compressibility profiles through the interior coefficients in Eqs.~\eqref{eq:eibi-f}-\eqref{eq:eibi-g}. The decreasing Love number therefore reflects a change in the internal structure as well as the smaller radius. Attributing this trend to a specific redistribution of matter would require comparing the interior profiles. The residuals have the same sign as \(\kappa\) and increase in magnitude across the five targets, as do the radius residuals. 

The absolute \(\Lambda\) values in Fig.~\ref{fig:lambda-mass} decrease by many orders of magnitude. Both changes in Eq.~\eqref{eq:tidal_def} act in the same direction: \(k_2\) falls while compactness rises as mass increases and radius decreases. The strong dependence on compactness accounts for this large range. On the residual scale, \(\delta_{k_2}\) and \(5\delta_R\) have the same sign at the sampled targets and reinforce one another. The radius term is the larger contribution to the linearized residual in Eq.~\eqref{eq:residual-linearized}. Thus the largest fractional EiBI response occurs toward high mass even though the absolute deformability is much smaller there.

The outermost positive and negative couplings give approximately antisymmetric high-mass residuals about GR. This is consistent with the leading weak-coupling terms in Eqs.~\eqref{eq:background-gr-limit} and~\eqref{eq:tidal-gr-limit}. The growth with mass is also consistent with the larger local deformation parameters \(8\pi\kappa\epshat\) and \(8\pi\kappa\phat\) at higher central density and pressure. 

The refined GR turning point is \(M=1.41661\,\Msun\) and \(\rho_c=2.31\times 10^{10}\,\mathrm{g\,cm^{-3}}\). Across the five sampled couplings, the turning mass increases monotonically with \(\kappa\). Its variation across the envelope family is below the displayed precision. Compared with the visible envelope dependence of radius and Love number discussed next, this weak dependence supports a distinction between a core-controlled mass extremum and surface-sensitive observables. Table~\ref{tab:turning-points} gives the baseline values and full logarithmic density bracket widths. The displayed density is the best sampled point within that bracket. 

\section{Sensitivity to the Regulated Surface Model}
\label{sec:surface-systematics}

We next hold mass and coupling fixed and vary the low-density completion. All nine envelopes are solved for every such pair. Let \(e\) label an envelope and \(e_{\rm ref}\) the baseline choice. For a positive observable \(X\), define the fractional half range
\begin{equation}
 \mathcal H_X(M,\kappa)=
 \frac{\max_e X(M,\kappa;e)-\min_e X(M,\kappa;e)}
 {2X(M,\kappa;e_{\rm ref})}.
 \label{eq:envelope-half-range}
\end{equation}
Over the sampled masses and couplings, the maximum fractional half ranges are \(0.0186\) in radius, \(0.0857\) in Love number, and \(3.02\times 10^{-5}\) in deformability.

\begin{figure}[!tbp]
\centering
\includegraphics[width=\columnwidth]{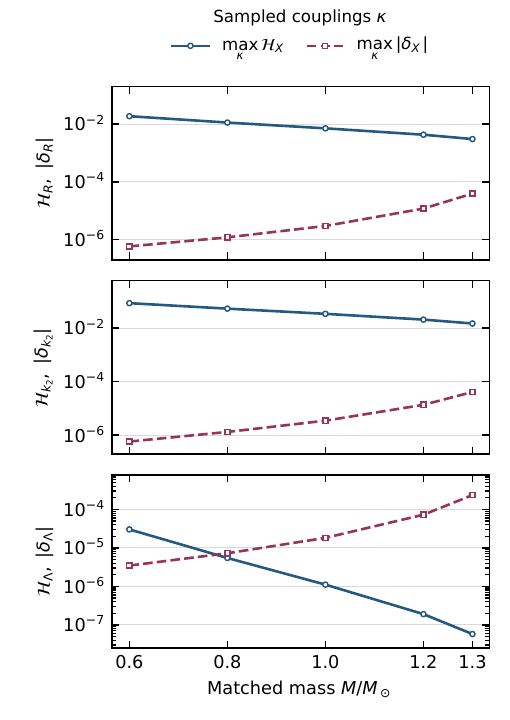}
\caption{Surface model sensitivity compared with the baseline envelope EiBI-GR separation at the five fixed-mass targets. Circles on solid lines show the largest \(\mathcal H_X\) over the sampled couplings, and squares on dashed lines show the largest absolute \(\delta_X\). From top to bottom, the panels give \(R\), \(k_2\), and \(\Lambda\) on logarithmic vertical axes. The envelope scale decreases while the gravity residual grows. }
\label{fig:surface-systematics}
\end{figure}

Figure~\ref{fig:surface-systematics} compares the maximum envelope half range with the largest absolute baseline envelope gravity residual at each mass. All three half ranges decrease across the targets, whereas the gravity residuals increase. The envelope scale remains larger for radius and Love number at all five masses. For deformability, the ordering reverses between the first two targets. The envelope scale is larger at \(0.6\,\Msun\) but smaller than the largest residual, attained at an outer sampled coupling, at the four higher masses. Table~\ref{tab:envelope-systematics} lists all cases. 

For two envelopes evaluated at the same fixed mass and coupling, the pairwise changes in \(k_2\) and \(R^5\) generally oppose one another. The fixed-mass scaling \(\Lambda\propto k_2R^5\) gives
\begin{equation}
 \Delta\ln\Lambda=\Delta\ln k_2+5\Delta\ln R.
 \label{eq:envelope-lambda-identity}
\end{equation}
This identity holds for any pair of envelope models. In our solutions, a larger envelope radius comes with a smaller Love number, so the two terms largely cancel and
$\Lambda$ changes much less than $k_2$ or $R^5$.

We also need the envelope dependence of the gravity residual itself. For envelope \(e\) we write
\begin{equation}
 \delta_X^{(e)}(M,\kappa)=
 \frac{X(M,\kappa;e)-X(M,0;e)}{X(M,0;e)}.
 \label{eq:envelope-gravity-residual}
\end{equation}
A common envelope effect can cancel between EiBI and GR, so \(\mathcal H_X\) does not determine the envelope dependence of this residual. We measure it separately through
\begin{equation}
 D_X=\frac{\max_e\delta_X^{(e)}-\min_e\delta_X^{(e)}}2.
 \label{eq:differential-envelope-spread}
\end{equation}
Here \(D_X\) is half the range of the fractional EiBI-GR residuals across the envelopes. At \(\kappa=0\), \(D_X=0\) by identity.

In double precision, $D_R$ and $D_{k_2}$ pass the $1\%$ refinement test at all nonzero couplings, but the much smaller $D_{\Lambda}$ does
not.  The final \(D_\Lambda\) values come from the high-precision calculations described below. 

\begin{figure}[!tbp]
\centering
\includegraphics[width=\columnwidth]{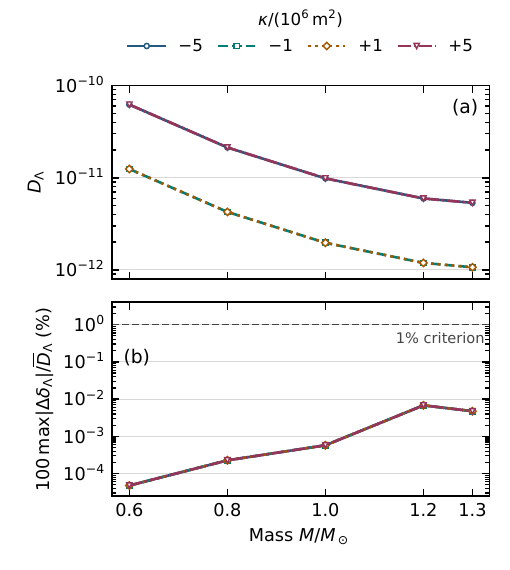}
\caption{Resolved differential deformability spread over the nine envelopes. Panel (a) gives the final high-precision \(D_\Lambda\) at each sampled mass and nonzero coupling. Panel (b) shows the largest per-envelope residual change divided by the mean spread, \(100E_{\rm pair}\), across the refinement levels of Appendix~\ref{app:hp}. The dashed line marks \(1\%\). Lines joining masses are visual guides.}
\label{fig:hp-dlambda}
\end{figure}

A high-precision calculation (Appendix~\ref{app:hp}) resolves this spread at all five masses and four nonzero couplings, comparing each EiBI model with GR at the same mass and envelope. \(D_\Lambda\) ranges from \(1.067\times 10^{-12}\) to \(6.208\times 10^{-11}\) (Fig.~\ref{fig:hp-dlambda}). Refining the integration, EOS table, precision and starting radius changes \(D_\Lambda\) by at most \(2.876\times 10^{-4}\%\) and the individual residuals by at most \(6.802\times 10^{-3}\%\) of \(D_\Lambda\).

A separate GR calculation with the bare Chandrasekhar EOS places these envelope variations in context. Table~\ref{tab:bare-gr} compares the bare model with the baseline regularized envelope at the same mass. Regularization increases the radius and decreases \(k_2\), and their effects largely cancel in \(\Lambda\). 
These offsets show the effects of the low-density regularization. The bare calculation is a GR benchmark.

\section{Physical Interpretation, Limitations, and Outlook}
\label{sec:limitations}

\subsection{Physical surface value at finite coupling}

At nonzero coupling, the equality \(y_F(r_s)=y_H^{(R)}(R)\) follows from the zero-density limit when \(1<\Gamma_{\rm env}<3/2\). Section~\ref{sec:tides} and Appendix~\ref{app:surface} establish it under the assumptions stated there.

Kim studied the local surface problem of the curvature singularity of an EiBI polytrope and its geodesic-deviation force, rather than the response to an externally applied quadrupole tide \cite{kim2014}. For positive \(\kappa\), he proposed that curvature-induced relative motion adds a term proportional to \(\varrho^{3/2}\) to the polytropic pressure, where \(\varrho\) is the energy density in his notation. The \(3/2\) scaling follows after choosing the characteristic length \(L_K=Rv_F/c\), where \(v_F\) is the Fermi velocity. This estimate uses \(v_F\propto\varrho^{1/3}\) and yields the marginal exponent for finite background curvature in that analysis. The length scale is heuristic and the computation is a local expansion near the surface. It does not give the surface derivative needed for \(k_2\) and \(\Lambda\).

At Kim's marginal exponent \(\Gamma_{\rm env}=3/2\), the term \(r\chi'/(1+\chi)\) keeps a finite EOS-dependent limit, so the auxiliary and physical surface values differ. 
Our choice \(1<\Gamma_{\rm env}<3/2\) removes this term and gives \(y_H^{(R)}(R)=y_F(r_s)\) at \(h=0\). Kim's work offers a possible microscopic motivation for
modified surface pressure. Our \(C^2\) envelope instead gives a closed tidal boundary condition used unchanged in both GR and EiBI.

Previous tidal calculations in EiBI gravity for neutron stars \cite{sham2014,prasetyo2021} match the auxiliary logarithmic derivative directly to the exterior. Our analysis shows that this agrees with the physical tidal response only if the low-density EOS satisfies the condition in Eq.~\eqref{eq:surface-equality}. Whether realistic neutron star crust EOSs do so deserves a dedicated check. 

\subsection{Interpreting the residuals and envelope response}

The residuals in Figs.~\ref{fig:mass-radius}-\ref{fig:lambda-mass} extend the known white dwarf response to a relativistic EiBI calculation with a specified regular surface. Their smooth weak-coupling ordering is compatible with the existing field equations. A controlled small-\(\kappa\) expansion at fixed mass, checked against additional resolved couplings, would be needed to turn the approximate antisymmetry into a quantitative linear response prediction.

The decrease of the envelope spread with mass in Fig.~\ref{fig:surface-systematics} has a simple hydrostatic explanation. Consider a thin power-law layer with nearly constant surface gravity \(g_s\). Hydrostatic balance gives \(dP/dr\simeq-\rho g_s\). For \(P=K\rho^{\Gamma_{\rm env}}\), the approximate thickness below \(\rho_{\rm tr}\), where the pressure is \(P_{\rm tr}\), is
\begin{equation}
 \Delta R_{\rm env}\simeq
 \frac{\Gamma_{\rm env}}{\Gamma_{\rm env}-1}
 \frac{P_{\rm tr}}{\rho_{\rm tr}g_s}.
 \label{eq:thin-envelope-thickness}
\end{equation}
The inverse dependence on \(g_s\) explains why a pressure layer change has less radial extent in a higher-gravity star. This supports the declining radius sensitivity toward higher mass. The actual relativistic EOS contains a one-decade blend as well as the analytic surface branch, so this estimate is only a rough guide to the half ranges.

The paired cancellation in Eq.~\eqref{eq:envelope-lambda-identity} suggests that moving the low-density surface changes the radius normalization more than it changes the dimensional quadrupole response. Testing that mechanism requires profile-resolved comparisons of the enclosed mass and tidal perturbation through the core, blend, and envelope. Establishing broader insensitivity would also require varying the blend width and function and using physically motivated atmosphere models.

In the Newtonian Poisson convention
\begin{equation}
 \nabla^2\Phi=4\pi G\rho+\frac{\kappa'}{4}\nabla^2\rho,
 \qquad \kappa'=8\pi G\kappa,
 \label{eq:kappa-conversion}
\end{equation}
the outer grid value \(|\kappa|=5\times10^6\,\mathrm{m^2}\) corresponds to \(|\kappa'|=8.39\times10^{-3}\,\mathrm{m^5\,kg^{-1}\,s^{-2}}\). Solar models give a much weaker scale, \(|\kappa'|\lesssim3\times10^5\) in these SI units \cite{casanellas2012}. A neutron star size argument gives a positive coupling bound of order \(10^{-2}\) \cite{avelino2012}, whereas a classical nuclear estimate gives \(|\kappa'|\lesssim10^{-3}\) \cite{avelino2012nuclear}. The latter assumes that gravity is subdominant in nuclear processes and that a continuous classical fluid description applies. Our outer couplings exceed that estimate. 

In Newtonian hydrostatic balance, Eq.~\eqref{eq:kappa-conversion} can be written using
\begin{equation}
 P_{\rm eff}^{\rm N}=P+\frac{\kappa'}8\rho^2
                  =P+\pi G\kappa\rho^2.
 \label{eq:newtonian-effective-pressure}
\end{equation}
In this Newtonian picture, a positive coupling adds pressure support. The exact mapping to Einstein equations for the auxiliary metric involves a nonlinear effective energy density and pressure \cite{delsate2012}, derived in Appendix~\ref{app:effective-fluid}.

\subsection{Scope and path to observational studies}

These predictions apply to a nonrotating, unmagnetized, zero-temperature barotropic perfect fluid under a static quadrupole tide. The core omits Coulomb and exchange-correlation corrections, reaction thresholds, and detailed composition. 
The envelope is a simple model with a fixed blend and does not contain opacity, radiative transfer, composition gradients or thermal stratification. 
Crystallization, elastic shear, and frequency-dependent tides are also omitted. The reported turning points are formal equilibrium sequence extrema. We do not compute
radial modes, so the stability of these models is not tested. 

For an eccentric binary star system, the static response gives the stellar structure factor in one component's classical equilibrium tide contribution to the slow rotation of the orbital ellipse (apsidal motion). At fixed component mass and fixed orbital elements, that individual term contains the dimensional tidal response \(k_2R^5\), multiplied by mass ratio, separation, and eccentricity factors \cite{sterne1939}. Because \(\Lambda\propto k_2R^5\) at fixed mass, \(\delta_\Lambda\) is the fractional change of this one term. 

Observational inference would require the dimensional response of each component, weighted by its baseline share of the total precession and combined with the point-mass and rotational terms. We do not compute this sum or a waveform, so we do not provide a detectability forecast or bound on the coupling.

\section{Conclusions}
\label{sec:conclusions}

We have formulated and calculated the static quadrupole tidal response of zero-temperature white dwarfs in relativistic EiBI gravity with a regulated surface. The physical surface derivative follows from differentiating the auxiliary-to-physical metric map. For \(P_{\rm env}=K\rho^{\Gamma_{\rm env}}\), the correction vanishes when \(1<\Gamma_{\rm env}<3/2\), giving the zero-density limit \(y_H^{(R)}(R)=y_F(r_s)\). The bare Chandrasekhar \(5/3\) edge does not have this limit. Cancellation free evaluation of the weak-coupling interior equations and low-compactness exterior formula then makes the small gravity residuals numerically accessible.

For the baseline envelope, over the five fixed-mass targets and sampled exploratory couplings, the largest absolute fractional EiBI-GR shifts are \(3.89\times 10^{-5}\) in radius, \(4.09\times 10^{-5}\) in \(k_2\), and \(2.36\times 10^{-4}\) in \(\Lambda\). Their magnitudes increase across the sampled masses, even as the absolute deformability decreases. The larger deformability residual follows from reinforcing Love number and fifth-power radius contributions.

Within the nine regulated envelopes, the largest fractional envelope half range for radius and Love number exceeds the largest absolute baseline envelope EiBI-GR residual at each target mass. For \(\Lambda\), the half range decreases across the five targets and exceeds that residual only at \(0.6\,\Msun\). Envelope-induced changes in \(k_2\) and \(R^5\) partly cancel, leaving the combined tidal response much less sensitive than either factor within this family. The absolute half ranges do not show how the EiBI-GR difference depends on the envelope. A high-precision calculation resolves its differential spread \(D_\Lambda\) at all 20 nonzero mass-coupling points under the stated refinement tests. 
This spread is small but the physical choice of the envelope remains open.  
The work provides tidal predictions for the chosen surface model and an explicit surface matching criterion. Establishing robustness beyond this envelope family and drawing observational constraints require the additional stellar and binary tests identified above.

\begin{acknowledgments}
This work makes use of Python packages. 
Claude (Anthropic) was used for help in LaTeX formatting and the uniform redrawing of some figures from our own data. 
All scientific contents are our own and we take full responsibility for these. 

\end{acknowledgments}

\section*{Code and Data Availability}

The source code and numerical data supporting this study will be made available from the corresponding author upon reasonable request.

\appendix

\section{Effective Fluid and Perturbation Map}
\label{app:effective-fluid}

The background and tidal equations follow from the effective fluid form of the auxiliary Einstein equations \cite{delsate2012}. Define \(R^\mu{}_\nu(q)=q^{\mu\alpha}R_{\alpha\nu}(q)\), with indices raised by the auxiliary metric. Equation~\eqref{eq:eibi-field-equations} gives
\begin{equation}
 R^\mu{}_\nu(q)=\frac{\delta^\mu{}_\nu-q^{\mu\alpha}g_{\alpha\nu}}{\kappa}.
 \label{eq:effective-ricci}
\end{equation}
In a frame adapted to the static fluid, \(q^{-1}g\) has temporal eigenvalue \(a/b^3=1+A_2\) and three spatial eigenvalues \(1/(ab)=1+A_1\). Taking the trace and forming the Einstein tensor yields
\begin{align}
 G^t{}_t(q)&=\frac{3A_1-A_2}{2\kappa}=-8\pi\epsilon_{\rm eff},\notag\\
 G^i{}_i(q)&=\frac{A_1+A_2}{2\kappa}=8\pi p_{\rm eff},\notag\\
 \epsilon_{\rm eff}&=\frac{A_2-3A_1}{16\pi\kappa},\qquad
 p_{\rm eff}=\frac{A_1+A_2}{16\pi\kappa}.
 \label{eq:effective-fluid}
\end{align}
There is no sum over \(i\). Here \(\epsilon_{\rm eff}\) and \(p_{\rm eff}\) are the geometrized energy density and pressure, both with units of \(\mathrm{m}^{-2}\). The auxiliary mass and lapse equations are consequently
\begin{equation}
 \mu'=\frac{4\pi r^2\epsilon_{\rm eff}}{\RsunG},\qquad
 \frac{\beta'}2=\frac{\RsunG\mu/r^2+4\pi r p_{\rm eff}}{D}.
 \label{eq:effective-background}
\end{equation}
These recover the mass equation~\eqref{eq:eibi-mass} and the lapse derivative in Eq.~\eqref{eq:tidal-intermediates}. Physical hydrostatic conservation gives \(\nu'=-2h'\), while differentiating the lapse map gives \(\beta'=-2(1+\chi)h'\), which recovers Eq.~\eqref{eq:eibi-enthalpy}.

For the static even-parity perturbation, angular fluid conservation fixes the Eulerian pressure amplitude,
\begin{equation}
 \delta P=-\frac{\epsilon+P}{2}H,\qquad
 \delta\epsilon=\frac{\delta P}{c_s^2}.
 \label{eq:static-fluid-amplitude}
\end{equation}
The common spherical harmonic is suppressed. Linearizing the temporal metric map in the same static fluid frame gives
\begin{align}
 H&=F+\delta\ln a-3\delta\ln b,\notag\\
 \delta\ln a&=\frac{4\pi\kappa\,\delta\widehat\epsilon}{a^2},
 \qquad
 \delta\ln b=-\frac{4\pi\kappa\,\delta\widehat P}{b^2}.
 \label{eq:linearized-lapse-map}
\end{align}
Substitution of Eq.~\eqref{eq:static-fluid-amplitude} gives \(H=F-\chi H\), hence \(H/F=(1+\chi)^{-1}\). This agrees with Eq.~(124) of Ref.~\cite{prasetyo2021}, written there in a different form. Its radial derivative and the circumferential radius chain rule give Eqs.~\eqref{eq:amplitude-map} and~\eqref{eq:physical-y-map}. This construction assumes a barotropic perfect fluid and a static tide.

\section{Cancellation-Free Algebra}
\label{app:stable-algebra}

To avoid cancellation in \(A_1\) and \(A_2\) near \(\kappa=0\) (Sec.~\ref{sec:numerics}), we define
\begin{equation}
 u=8\pi\kappa\epshat,\qquad v=8\pi\kappa\phat,
 \qquad a^2=1+u,\qquad b^2=1-v,
 \label{eq:uv-definitions}
\end{equation}
and the kernels
\begin{align}
 L(x)&=\frac{\ln(1+x)}{x},
 \qquad E(x)=\frac{e^x-1}{x},\notag\\
 \operatorname{sinhc}(x)&=\frac{\sinh x}{x},
 \label{eq:regular-kernels}
\end{align}
with their values at the origin defined by analytic continuation. If \(B=-2\ln b\) and \(d=\ln(a/b)\), then
\begin{align}
 \frac{B}{\kappa}&=(8\pi\phat)L(-v),
 \label{eq:B-over-kappa}\\
 \frac{d}{\kappa}&=4\pi\epshat L(u)+4\pi\phat L(-v).
 \label{eq:d-over-kappa}
\end{align}
The dimensionless exponents \(\zeta_-=B-d\) and \(\zeta_+=B+d\) give the exact quotient forms
\begin{equation}
 \frac{A_1}{\kappa}=\frac{\zeta_-}{\kappa}E(\zeta_-),
 \qquad
 \frac{A_2}{\kappa}=\frac{\zeta_+}{\kappa}E(\zeta_+).
 \label{eq:A-quotients}
\end{equation}
Regrouping these quotients gives the combinations in the background equations:
\begin{align}
 \frac{\mathcal A}{\kappa}={}&2\frac{B}{\kappa}E(B)
 +e^B d\frac{d}{\kappa}\operatorname{sinhc}^2(d/2),
 \label{eq:A-sum-stable}\\
 \frac{A_2-3A_1}{\kappa}={}&
 2e^B\frac{d}{\kappa}\operatorname{sinhc}(d)
 -2\frac{\zeta_-}{\kappa}E(\zeta_-).
 \label{eq:mass-stable}
\end{align}
Since \(a/b^3=e^{\zeta_+}\), we evaluate the tidal combination as
\begin{equation}
 \frac{\mathcal Q}{\kappa}
 =\frac{\zeta_+}{\kappa}E(\zeta_+)
 +\frac{e^{\zeta_+}}{1+\chi}\frac{\chi}{\kappa}.
 \label{eq:Q-stable}
\end{equation}
The kernels remain finite as \(\kappa\rightarrow0\), so these expressions have smooth GR limits.

Near the center we also write \(f=2/r+\widetilde f\) and \(g=-\ell(\ell+1)/r^2+\widetilde g\), so that the \(1/r\) terms in Eq.~\eqref{eq:riccati} cancel analytically at \(y_F=\ell\).

\section{Center Expansion}
\label{app:center}

The regular solution can be initialized without evaluating the singular coordinate equations at \(r=0\). A local power series (Frobenius expansion) gives the state at a small positive radius. With the subscript \(c\) denoting central values,
\begin{align}
 h&=h_c+h_2r^2+O(r^4),&
 \mu&=m_3r^3+O(r^5),\notag\\
 y_F&=\ell+y_2r^2+O(r^4),&
 f&=\frac{2}{r}+f_1r+O(r^3),\notag\\
 g&=-\frac{\ell(\ell+1)}{r^2}+g_0+O(r^2).
 \label{eq:center-series}
\end{align}
Substitution into Eq.~\eqref{eq:riccati} yields
\begin{equation}
 y_2=-\frac{\ell f_1+g_0}{2\ell+3}.
 \label{eq:center-y2}
\end{equation}
For the exact GR branch, the coefficients are
\begin{align}
 m_3&=\frac{4\pi\epshat_c}{3\RsunG}~,~~
 h_2=-\frac{\RsunG m_3}{2}-2\pi\phat_c,
 \label{eq:center-gr-background}\\
 f_1&=2\RsunG\,m_3+4\pi(\phat_c-\epshat_c),
 \notag\\
 g_0&=-2\ell(\ell+1)\RsunG\,m_3\notag\\
 &\quad+4\pi(5\epshat_c+9\phat_c+\Sigmahat_c).
 \label{eq:center-gr-tides}
\end{align}
For finite nonzero \(\kappa\), every unmarked algebraic factor below is evaluated at the center:
\begin{align}
 m_3&=\frac{(A_2-3A_1)/\kappa}{12\RsunG},
 \label{eq:center-eibi-mass}\\
 h_2&=-\frac{\mathcal A/\kappa}{8(1+\chi)}
 -\frac{\RsunG\,m_3}{2(1+\chi)},
 \label{eq:center-eibi-enthalpy}\\
 f_1&=\frac{A_1}{\kappa}+2\RsunG\,m_3,
 \label{eq:center-eibi-f}\\
 g_0&=2\left[\frac{A_2}{\kappa}
 +\frac{a}{b^3}\frac{\chi/\kappa}{1+\chi}\right]
 -2\ell(\ell+1)\RsunG\,m_3.
 \label{eq:center-eibi-g}
\end{align}
The identities in Appendix~\ref{app:stable-algebra} evaluate these finite \(\kappa\) coefficients and recover their GR limits continuously. The resulting initial state is
\begin{align}
 h(r_0)&=h_c+h_2r_0^2,
 \qquad \mu(r_0)=m_3r_0^3,\notag\\
 y_F(r_0)&=\ell+y_2r_0^2,
 \label{eq:center-launch-state}
\end{align}
with \(r_0=5\times10^{-4}\,\mathrm m\) in the baseline integrations. The omitted terms are \(O(r_0^4)\), \(O(r_0^5)\), and \(O(r_0^4)\), respectively. The integration therefore starts from regular finite radius data.

\section{Surface Limit}
\label{app:surface}

To derive the physical tidal matching condition, consider finite nonzero \(\kappa\) and assume \(K>0\), \(e_0>0\), and \(1<\Gamma_{\rm env}<2\) in Eq.~\eqref{eq:surface-eos}. Direct differentiation gives
\begin{equation}
 c_s^2=
 \frac{\Gamma_{\rm env}K\rho^{\Gamma_{\rm env}-1}}
 {e_0+\Gamma_{\rm env}K\rho^{\Gamma_{\rm env}-1}/
 (\Gamma_{\rm env}-1)}
 =O(\rho^{\Gamma_{\rm env}-1}).
 \label{eq:surface-sound-speed}
\end{equation}
The ratio in Eq.~\eqref{eq:sigma-definition} then gives
\begin{align}
 \Sigma&=\frac{e_0^2}{\Gamma_{\rm env}K}
 \rho^{2-\Gamma_{\rm env}}
 \left[1+O(\rho^{\Gamma_{\rm env}-1})\right],\notag\\
 \chi&=O(\rho^{2-\Gamma_{\rm env}}).
 \label{eq:surface-sigma-chi}
\end{align}
Expanding the enthalpy of the same surface EOS gives
\begin{equation}
 h=\frac{\Gamma_{\rm env}K}
 {(\Gamma_{\rm env}-1)e_0}
 \rho^{\Gamma_{\rm env}-1}
 +O(\rho^{2\Gamma_{\rm env}-2}).
 \label{eq:surface-enthalpy-order}
\end{equation}

At a nondegenerate surface with \(M>0\) and \(D(r_s)>0\), Eq.~\eqref{eq:eibi-enthalpy} has a finite negative one-sided limit,
\begin{equation}
 \left.\frac{dh}{dr}\right|_{r_s^-}
 =-\frac{\RsunG\mu(r_s)}{r_s^2D(r_s)}.
 \label{eq:surface-enthalpy-gradient}
\end{equation}
Combining this limit with Eq.~\eqref{eq:surface-enthalpy-order} gives
\begin{align}
 \rho'&=O(\rho^{2-\Gamma_{\rm env}}),
 \qquad
 \chi_{,\rho}=O(\rho^{1-\Gamma_{\rm env}}),\notag\\
 r\chi'&=O(\rho^{3-2\Gamma_{\rm env}}).
 \label{eq:surface-derivative-proof}
\end{align}
The leading coefficient can also be retained. Writing \(h'_s=\lim_{r\to r_s^-}h'\) and
\begin{equation}
 B_\chi=\frac{2\pi\kappa G e_0^2}{c^4\Gamma_{\rm env}K},
\end{equation}
we obtain
\begin{equation}
 r\chi'\sim
 \frac{r_s B_\chi(2-\Gamma_{\rm env})e_0h'_s}
 {\Gamma_{\rm env}K}\,
 \rho^{3-2\Gamma_{\rm env}}.
 \label{eq:surface-leading-coefficient}
\end{equation}
For finite nonzero \(\kappa\), \(K,e_0>0\), \(r_s>0\), and finite nonzero \(h'_s\), this coefficient is nonzero. Thus the orders in Eq.~\eqref{eq:surface-derivative-proof} give a finite nonzero limit at \(\Gamma_{\rm env}=3/2\) and a divergent magnitude for \(3/2<\Gamma_{\rm env}<2\). Degenerate surfaces or different EOS asymptotics require separate analysis.

The metric derivatives satisfy \(a'/a=O(\rho^{2-\Gamma_{\rm env}})\) and \(b'/b=O(\rho)\), while the deformation factors approach unity. Thus the circumferential-to-auxiliary radius ratio tends to unity and Eq.~\eqref{eq:physical-radius-derivative} gives \(dR_{\rm circ}/dr\to1\). Together, Eqs.~\eqref{eq:physical-y-map} and \eqref{eq:surface-derivative-proof} give three regimes:
\begin{align}
 1<\Gamma_{\rm env}<\frac32
 &:\quad r\chi'\longrightarrow0,\notag\\
 \Gamma_{\rm env}=\frac32
 &:\quad r\chi'\longrightarrow\hbox{finite nonzero constant},\notag\\
 \frac32<\Gamma_{\rm env}<2
 &:\quad |r\chi'|\longrightarrow\infty.
 \label{eq:surface-three-regimes}
\end{align}
The first case gives Eq.~\eqref{eq:surface-equality}. For the bare Chandrasekhar exponent \(5/3\), the finite-coupling physical map instead gives \(r\chi'=O(\rho^{-1/3})\). Surfaces with \(\Gamma_{\rm env}\geq2\) require a separate asymptotic analysis. At \(\kappa=0\), \(\chi=0\) identically and the two tidal variables coincide without this finite-coupling argument.

\section{Validation and Numerical Accuracy Details}
\label{app:validation}

The Radau checks use relative tolerance \(10^{-12}\), absolute tolerance \(10^{-14}\), maximum step \(5\,\mathrm{km}\) and surface tolerance \(10^{-13}\). The tighter solutions of Sec.~\ref{sec:validation} use relative tolerance \(2.5\times10^{-14}\), absolute tolerance \(2.5\times10^{-16}\), maximum step \(1250\,\mathrm m\) and starting radius \(2.5\times10^{-4}\,\mathrm m\). The same settings are used to locate each mass maximum, by bisection until the bracket is narrower than both \(\Delta\log_{10}\rho_c\le5\times10^{-3}\) and the successive best-mass relative change  \(\frac{|M_{\rm best}^{(n)}-M_{\rm best}^{(n-1)}|}{|M_{\rm best}^{(n)}|}\leq10^{-7}\), and to repeat the bare EOS benchmark of Table~\ref{tab:bare-gr}, which then changes \(R\), \(k_2\) and \(\Lambda\) by at most \(4.522\times10^{-13}\).

Refining the EOS table from 256 to 512 and 1024 points per density decade, for all nine envelopes at \(M=0.6\) and \(1.3\,M_\odot\) and \(\kappa=0\) and \(+5\times10^6\,\mathrm{m^2}\), changes the observables, \(\mathcal H_X\), \(\delta_X\), \(D_R\) and \(D_{k_2}\) by at most \(5.393\times10^{-4}\) (Table~\ref{tab:eos-refinement}). The much smaller \(D_\Lambda\) does not converge in double precision. At \(M=1.3\,M_\odot\) and \(\kappa=+5\times10^6\,\mathrm{m^2}\), reducing the radial step near the EOS joins and the surface from \(31.25\) to \(7.8125\,\mathrm m\) changes it by \(6.162\%\) in symmetric relative difference. We therefore compute \(D_\Lambda\) at high precision.

\subsection{High-precision differential deformability calculation}
\label{app:hp}

The precision calculation uses the same physical EOS, metric map and vacuum matching. 
Only the EOS representation, the independent variable and the arithmetic precision change. The input EOS constants are 
taken unchanged from the double-precision calculation and all derived EOS quantities are re-calculated at the working precision. The
values of $G$, $c$ and $M_{\odot}$ are unchanged. 

The first law and the surface normalization in Eq.~\eqref{eq:surface-e0} give
\begin{align}
 h&=\ln\frac{\epsilon+P}{e_0\rho},\notag\\
 h_{\rm core}&=\ln[1+\eta(\sqrt{1+x^2}-1)]+h_{\rm off},\notag\\
 \eta&=\frac{m_e}{\mu_e m_u},\qquad h_{\rm off}=\ln\frac{c^2}{e_0}.
 \label{eq:hp-enthalpy-offset}
\end{align}
Although the physical core is unchanged, its enthalpy offset depends on the envelope. Omitting this offset would compare different physical states at the same numerical enthalpy. With \(z=[e^{h-h_{\rm off}}-1]/\eta\), the analytic core inverse is \(\rho=\rho_\star[z(z+2)]^{3/2}\). In the blend, we represent the small specific energy correction \(\omega=\epsilon/\rho-c^2\), whose first law target is \(d\omega/d\ln\rho=P/\rho\), using piecewise polynomial integration. Direct enthalpy inversion preserves the normalization and both joins. The analytic pressure is unchanged and the numerical energy representation is checked separately against higher-precision quadrature.

Write \(E_q=\epsilon_{\rm eff}\), \(P_q=p_{\rm eff}\) for the geometrized effective-fluid variables in Eq.~\eqref{eq:effective-fluid}, and set \(W=\epshat+\phat\), \(J=1+\chi\). Cancellation-free evaluation uses
\begin{align}
 P_q&=\frac{\phat}{b^2}
       +\frac{4\pi\kappa W^2}{(a+b)^2ab^3},\notag\\
 E_q&=\frac{W}{ab^3}-P_q,\qquad
 \Sigma_q=\frac{1}{J}\frac{dE_q}{dh}.
 \label{eq:hp-apparent-fluid}
\end{align}
Here \(J=(dP_q/dh)/(E_q+P_q)\), so \(\Sigma_q=(E_q+P_q)\,dE_q/dP_q\). Here \(\Sigma_q\) describes the pressure-density response of the transformed matter source in the auxiliary metric equations. At nonzero coupling, it generally differs from the corresponding response of the physical stellar matter. These expressions have continuous GR limits and use \(d\epshat/dh=\Sigmahat\), \(d\phat/dh=W\). We checked these identities numerically.

For the high-precision integration we use the dimensionless variables
\begin{align}
 t_h&=1-h/h_c,\qquad u_h=r^2/L^2,\notag\\
 v_h&=\RsunG\mu/r,\qquad w=y_F-2,\notag\\
 L&=10^6\,\mathrm m,\qquad D=1-2v_h,\notag\\
 B_h&=v_h+4\pi L^2u_hP_q,\qquad A_h=h_cDJ/B_h.
 \label{eq:hp-scaled-variables}
\end{align}
The quadrupole equations become
\begin{align}
 \frac{du_h}{dt_h}&=2A_hu_h,\qquad
 \frac{dv_h}{dt_h}=A_h(4\pi L^2u_hE_q-v_h),\notag\\
 \frac{dw}{dt_h}&=A_h\left(\frac{T}{D}+\frac{4B_h^2}{D^2}\right),\notag\\
 T&=-w(w+5)+2v_h(2+w)^2\notag\\
 &\quad-4\pi L^2u_h[(2+w)(P_q-E_q)+5E_q+9P_q+\Sigma_q].
 \label{eq:hp-scaled-trajectory}
\end{align}
Here \(r\) remains the auxiliary radial coordinate and \(\mu\) the mass in solar units. This reformulation retains the small central correction \(w\) explicitly. The center coefficients in Appendix~\ref{app:center} give
\begin{align}
 t_{h,0}&=-h_2r_0^2/h_c,\notag\\
 (u_{h,0},v_{h,0},w_0)&=(r_0^2/L^2,\RsunG m_3r_0^2,y_2r_0^2).
 \label{eq:hp-center-launch}
\end{align}
We compute \(t_{h,0}\) directly. The linearized tidal equation still contains a coefficient approaching \(-5/(2t_h)\) near the center, so the first step is \(t_{h,0}/10\). We integrate with the embedded Runge-Kutta-Fehlberg 7(8) method at the working precision, ending steps exactly at both EOS joins and at the surface \(t_h=1\). There we impose vacuum and set \(R_{\rm circ}=r_s\) and \(y_H^{(R)}=y_F\).

The first level uses 60 decimal digits, EOS polynomials of degree 24 on eight blend pieces, \(r_0=1.25\times10^{-4}\,\mathrm m\) and relative tolerance \(10^{-20}\). Successive levels tighten the tolerance to \(10^{-22}\) and \(10^{-24}\), raise the polynomial degree to 32, raise the precision to 80 digits, and halve \(r_0\). At all refinement levels, the component absolute tolerances for \((u_h,v_h,w)\) are \((10^{-45},10^{-48},10^{-48})\), and the maximum step in \(t_h\) is \(0.01\). At every level the mass is matched to within \(10^{-24}\,M_\odot\). The calculation contains \(1350\) stellar models, with largest mass residual \(9.980\times 10^{-25}\,M_\odot\).

For two levels \(a,b\), put \(\overline D_\Lambda=(D_{\Lambda,a}+D_{\Lambda,b})/2\). Besides the symmetric spread change \(|D_{\Lambda,a}-D_{\Lambda,b}|/\overline D_\Lambda\), we test
\begin{equation}
 E_{\rm pair}=\frac{\max_e|\delta_{\Lambda,a}^{(e)}-
                          \delta_{\Lambda,b}^{(e)}|}{\overline D_\Lambda}.
 \label{eq:hp-pair-refinement}
\end{equation}
This second test ensures that the individual residuals, and not only their spread, are converged. Both measures change by less than \(1\%\) between all successive levels and at all 20 points. The reported values use the final level.

As a final check, we recomputed \(\Lambda\) from every stored surface solution in two ways, using the closed Love number formula and a direct exterior solution at 140 digits 
and found agreement. 

\raggedbottom
\onecolumngrid
\section{Selected Numerical Results}
\label{app:numerical-results}

This appendix collects the numerical results referred to in the text. Table~\ref{tab:gr-benchmarks} lists the tests of the calculation and the comparison with Perot and Chamel. Table~\ref{tab:turning-points} reports the baseline-envelope extrema with their density brackets. Table~\ref{tab:envelope-systematics} gives the complete sampled fixed-mass gravity residuals and absolute envelope half ranges.

\begingroup
\squeezetable
\begin{table}[H]
\caption{Numerical benchmarks and a separate literature comparison. The first group uses numerical reference values for the Newtonian \(n=3/2\) polytrope. The \(k_2\) convention is also checked against the rounded tabulation of Ref.~\cite{brooker1955}. Here \(\theta(\xi)\) is the Lane-Emden function, \(\xi_1\) its first zero, \(\theta'_1=\theta'(\xi_1)\), and \(\eta_2\) the surface Clairaut-Radau variable in Eq.~\eqref{eq:clairaut-radau-k2}. The second group compares the \(\kappa=0\) branch with a separate GR solver at \(\rho_c=10^9\,\mathrm{g\,cm^{-3}}\), using the same baseline EOS. The last group compares \(k_2\) with the fluid white dwarf fit of Perot and Chamel~\cite{perot2022}, whose EOS differs. Absolute differences are \(|X-X_{\rm ref}|\), and relative differences divide this by \(|X_{\rm ref}|\). Differences use unrounded values.}
\label{tab:gr-benchmarks}
\setlength{\tabcolsep}{3pt}
\begin{ruledtabular}
\begin{tabular}{lrrrlr}
Quantity / mass & This work & Reference & Difference & Convention & Tolerance \\
\hline
\multicolumn{6}{l}{\textit{Newtonian $n=3/2$ polytrope: Lane-Emden and Clairaut-Radau checks}} \\
$k_2(n=3/2)$ & $0.143279$ & $0.143279$ & $1.77\times 10^{-12}$ & relative & $1\times 10^{-10}$ \\
$-\xi_1^2\theta'_1$ & $2.71406$ & $2.71406$ & $2.53\times 10^{-13}$ & relative & $1\times 10^{-10}$ \\
$\xi_1$ & $3.65375$ & $3.65375$ & $1.58\times 10^{-15}$ & relative & $1\times 10^{-10}$ \\
$\eta_2$ & $1.88634$ & $1.88634$ & $8.11\times 10^{-13}$ & relative & $1\times 10^{-10}$ \\
\hline
\multicolumn{6}{l}{\textit{Independent GR implementation check: same EOS, $\kappa=0$}} \\
$k_2$ & $0.0438816$ & $0.0438816$ & $2.89\times 10^{-11}$ & relative & $1\times 10^{-5}$ \\
$M/M_\odot$ & $1.35898$ & $1.35898$ & $8.19\times 10^{-13}$ & relative & $1\times 10^{-7}$ \\
$R$ [m] & $2.46863\times 10^{6}$ & $2.46863\times 10^{6}$ & $1.63\times 10^{-12}$ & relative & $1\times 10^{-6}$ \\
$y_s$ & $1.59638$ & $1.59638$ & $6.74\times 10^{-12}$ & relative & $1\times 10^{-5}$ \\
$\Lambda$ & $8.24242\times 10^{13}$ & $8.24242\times 10^{13}$ & $4.11\times 10^{-11}$ & relative & $5\times 10^{-5}$ \\
\hline
\multicolumn{6}{l}{\textit{White dwarf comparison: Perot and Chamel (different EOS)}} \\
$k_2$ ($M=0.6\,M_\odot$) & $0.11533$ & $0.123223$ & $0.00789$ & absolute & --- \\
$k_2$ ($M=0.8\,M_\odot$) & $0.10322$ & $0.107278$ & $0.00406$ & absolute & --- \\
$k_2$ ($M=1.0\,M_\odot$) & $0.0876572$ & $0.0887$ & $0.00104$ & absolute & --- \\
\end{tabular}
\end{ruledtabular}
\end{table}
\endgroup

\begingroup
\squeezetable
\begin{table}[H]
\caption{Refined equilibrium sequence maxima for the baseline envelope, \(\Gamma_{\rm env}=4/3\) and \(\rho_{\rm tr}=10^3\,\mathrm{g\,cm^{-3}}\). \(M_{\rm tp}\) is the largest computed mass in the refined bracket and \(\rho_{c,\rm tp}\) its central density. The last column is the full width of the logarithmic density bracket, about \(0.9\%\) in density.}
\label{tab:turning-points}
\setlength{\tabcolsep}{3pt}
\begin{ruledtabular}
\begin{tabular}{rrrr}
$\kappa$ [m$^2$] & $M_{\rm tp}/M_\odot$ & $\rho_{c,\rm tp}$ [g cm$^{-3}$] & $\Delta\log_{10}\rho_c$ \\
\hline
$-5\times 10^{6}$ & $1.41643$ & $2.28\times 10^{10}$ & $0.00391$ \\
$-1\times 10^{6}$ & $1.41658$ & $2.31\times 10^{10}$ & $0.00391$ \\
$0$ & $1.41661$ & $2.31\times 10^{10}$ & $0.00391$ \\
$1\times 10^{6}$ & $1.41665$ & $2.32\times 10^{10}$ & $0.00391$ \\
$5\times 10^{6}$ & $1.4168$ & $2.35\times 10^{10}$ & $0.00391$ \\
\end{tabular}
\end{ruledtabular}
\end{table}
\endgroup

\begingroup
\squeezetable
\begin{table}[H]
\caption{Gravity response and envelope dependence at fixed mass. The baseline envelope is \(\Gamma_{\rm env}=4/3\), \(\rho_{\rm tr}=10^3\,\mathrm{g\,cm^{-3}}\). For \(X\in\{R,k_2,\Lambda\}\), \(\delta_X=[X(\kappa)-X(0)]/X(0)\) uses that same envelope in both models. The fractional envelope half range is \(\mathcal H_X=(\max_e X-\min_e X)/(2X_{\rm base})\) at the listed mass and coupling, over all nine envelopes. Both \(\delta_X\) and \(\mathcal H_X\) are dimensionless. The half ranges describe variation of the absolute observables.}
\label{tab:envelope-systematics}
\setlength{\tabcolsep}{3pt}
\begin{ruledtabular}
\begin{tabular}{rrrrrrrr}
$\kappa\,[\mathrm{m}^2]$ & $M/M_\odot$ & $\delta_R$ & $\delta_{k_2}$ & $\delta_\Lambda$ & $\mathcal H_R$ & $\mathcal H_{k_2}$ & $\mathcal H_\Lambda$ \\
\hline

$-5\times 10^{6}$ & $0.6$ & $-5.79\times 10^{-7}$ & $-5.77\times 10^{-7}$ & $-3.47\times 10^{-6}$ & $0.0186$ & $0.0857$ & $3.02\times 10^{-5}$ \\

$-5\times 10^{6}$ & $0.8$ & $-1.19\times 10^{-6}$ & $-1.32\times 10^{-6}$ & $-7.25\times 10^{-6}$ & $0.0112$ & $0.0534$ & $5.47\times 10^{-6}$ \\

$-5\times 10^{6}$ & $1$ & $-2.95\times 10^{-6}$ & $-3.48\times 10^{-6}$ & $-1.82\times 10^{-5}$ & $0.00707$ & $0.0342$ & $1.11\times 10^{-6}$ \\

$-5\times 10^{6}$ & $1.2$ & $-1.18\times 10^{-5}$ & $-1.36\times 10^{-5}$ & $-7.28\times 10^{-5}$ & $0.00425$ & $0.0208$ & $1.9\times 10^{-7}$ \\

$-5\times 10^{6}$ & $1.3$ & $-3.89\times 10^{-5}$ & $-4.09\times 10^{-5}$ & $-2.36\times 10^{-4}$ & $0.00301$ & $0.0149$ & $5.76\times 10^{-8}$ \\

$-1\times 10^{6}$ & $0.6$ & $-1.16\times 10^{-7}$ & $-1.15\times 10^{-7}$ & $-6.94\times 10^{-7}$ & $0.0186$ & $0.0857$ & $3.02\times 10^{-5}$ \\

$-1\times 10^{6}$ & $0.8$ & $-2.37\times 10^{-7}$ & $-2.64\times 10^{-7}$ & $-1.45\times 10^{-6}$ & $0.0112$ & $0.0534$ & $5.47\times 10^{-6}$ \\

$-1\times 10^{6}$ & $1$ & $-5.9\times 10^{-7}$ & $-6.97\times 10^{-7}$ & $-3.64\times 10^{-6}$ & $0.00707$ & $0.0342$ & $1.11\times 10^{-6}$ \\

$-1\times 10^{6}$ & $1.2$ & $-2.37\times 10^{-6}$ & $-2.72\times 10^{-6}$ & $-1.46\times 10^{-5}$ & $0.00425$ & $0.0208$ & $1.9\times 10^{-7}$ \\

$-1\times 10^{6}$ & $1.3$ & $-7.78\times 10^{-6}$ & $-8.18\times 10^{-6}$ & $-4.71\times 10^{-5}$ & $0.00301$ & $0.0149$ & $5.76\times 10^{-8}$ \\

$0$ & $0.6$ & $0$ & $0$ & $0$ & $0.0186$ & $0.0857$ & $3.02\times 10^{-5}$ \\

$0$ & $0.8$ & $0$ & $0$ & $0$ & $0.0112$ & $0.0534$ & $5.47\times 10^{-6}$ \\

$0$ & $1$ & $0$ & $0$ & $0$ & $0.00707$ & $0.0342$ & $1.11\times 10^{-6}$ \\

$0$ & $1.2$ & $0$ & $0$ & $0$ & $0.00425$ & $0.0208$ & $1.9\times 10^{-7}$ \\

$0$ & $1.3$ & $0$ & $0$ & $0$ & $0.00301$ & $0.0149$ & $5.76\times 10^{-8}$ \\

$1\times 10^{6}$ & $0.6$ & $1.16\times 10^{-7}$ & $1.16\times 10^{-7}$ & $6.94\times 10^{-7}$ & $0.0186$ & $0.0857$ & $3.02\times 10^{-5}$ \\

$1\times 10^{6}$ & $0.8$ & $2.37\times 10^{-7}$ & $2.64\times 10^{-7}$ & $1.45\times 10^{-6}$ & $0.0112$ & $0.0534$ & $5.47\times 10^{-6}$ \\

$1\times 10^{6}$ & $1$ & $5.9\times 10^{-7}$ & $6.97\times 10^{-7}$ & $3.64\times 10^{-6}$ & $0.00707$ & $0.0342$ & $1.11\times 10^{-6}$ \\

$1\times 10^{6}$ & $1.2$ & $2.37\times 10^{-6}$ & $2.72\times 10^{-6}$ & $1.46\times 10^{-5}$ & $0.00425$ & $0.0208$ & $1.9\times 10^{-7}$ \\

$1\times 10^{6}$ & $1.3$ & $7.78\times 10^{-6}$ & $8.18\times 10^{-6}$ & $4.71\times 10^{-5}$ & $0.00301$ & $0.0149$ & $5.76\times 10^{-8}$ \\

$5\times 10^{6}$ & $0.6$ & $5.79\times 10^{-7}$ & $5.77\times 10^{-7}$ & $3.47\times 10^{-6}$ & $0.0186$ & $0.0857$ & $3.02\times 10^{-5}$ \\

$5\times 10^{6}$ & $0.8$ & $1.19\times 10^{-6}$ & $1.32\times 10^{-6}$ & $7.25\times 10^{-6}$ & $0.0112$ & $0.0534$ & $5.47\times 10^{-6}$ \\

$5\times 10^{6}$ & $1$ & $2.95\times 10^{-6}$ & $3.48\times 10^{-6}$ & $1.82\times 10^{-5}$ & $0.00707$ & $0.0342$ & $1.11\times 10^{-6}$ \\

$5\times 10^{6}$ & $1.2$ & $1.18\times 10^{-5}$ & $1.36\times 10^{-5}$ & $7.28\times 10^{-5}$ & $0.00425$ & $0.0208$ & $1.9\times 10^{-7}$ \\

$5\times 10^{6}$ & $1.3$ & $3.89\times 10^{-5}$ & $4.09\times 10^{-5}$ & $2.36\times 10^{-4}$ & $0.00301$ & $0.0149$ & $5.76\times 10^{-8}$ \\
\end{tabular}
\end{ruledtabular}
\end{table}
\endgroup

\begingroup
\squeezetable
\begin{table}[H]
\caption{EOS table refinement in double precision. \(N\) is the number of interpolation points per decade of density. The first column gives the two resolutions compared. \(s\) is defined in Sec.~\ref{sec:validation}, with \(s(0,0)=0\). Each entry is the largest change over nine envelopes, masses \(0.6\) and \(1.3\,M_\odot\), and couplings \(0\) and \(+5\times10^6\,\mathrm{m^2}\). \(X\) is the absolute observable, \(\mathcal H_X\) its envelope half range, \(\delta_X\) its same-envelope gravity residual, and \(D_X\) the envelope half range of that residual. The \(\delta_X\) comparison includes each envelope separately. \(D_\Lambda\) is not converged at this precision. Figure~\ref{fig:hp-dlambda} shows the high-precision values.}
\label{tab:eos-refinement}
\centering
\setlength{\tabcolsep}{6pt}
\begin{tabular}{llrrrr}
\hline
$N_a\to N_b$ & Observable $X$ & $\max s(X)$ & $\max s(\mathcal H_X)$ & $\max s(\delta_X)$ & $\max s(D_X)$ \\
\hline
$256\!\to\!512$ & $R$ & $3.595\times10^{-12}$ & $1.859\times10^{-10}$ & $4.600\times10^{-7}$ & $1.823\times10^{-6}$ \\
$256\!\to\!512$ & $k_2$ & $1.011\times10^{-10}$ & $2.113\times10^{-9}$ & $2.502\times10^{-4}$ & $3.494\times10^{-4}$ \\
$256\!\to\!512$ & $\Lambda$ & $1.094\times10^{-10}$ & $5.736\times10^{-4}$ & $4.089\times10^{-5}$ & $1.397\times10^{0}$ \\
$512\!\to\!1024$ & $R$ & $7.152\times10^{-13}$ & $2.649\times10^{-11}$ & $2.229\times10^{-7}$ & $5.710\times10^{-6}$ \\
$512\!\to\!1024$ & $k_2$ & $2.917\times10^{-10}$ & $1.878\times10^{-9}$ & $4.993\times10^{-4}$ & $3.046\times10^{-3}$ \\
$512\!\to\!1024$ & $\Lambda$ & $2.925\times10^{-10}$ & $9.486\times10^{-5}$ & $8.473\times10^{-5}$ & $1.137\times10^{0}$ \\
$256\!\to\!1024$ & $R$ & $3.461\times10^{-12}$ & $1.898\times10^{-10}$ & $3.370\times10^{-7}$ & $7.130\times10^{-6}$ \\
$256\!\to\!1024$ & $k_2$ & $3.279\times10^{-10}$ & $1.782\times10^{-9}$ & $5.393\times10^{-4}$ & $2.697\times10^{-3}$ \\
$256\!\to\!1024$ & $\Lambda$ & $3.396\times10^{-10}$ & $4.788\times10^{-4}$ & $9.188\times10^{-5}$ & $1.504\times10^{0}$ \\
\hline
\end{tabular}
\end{table}
\endgroup

\begingroup
\squeezetable
\begin{table}[H]
\caption{Effect of the low-density regularization in GR (\(\kappa=0\)), at fixed mass and \(\mu_e=2\). Bare GR uses the Chandrasekhar EOS without an envelope modification. Regularized GR uses the baseline envelope, \(\Gamma_{\rm env}=4/3\) and \(\rho_{\rm tr}=10^3\,\mathrm{g\,cm^{-3}}\). Radius is in km, while \(k_2\) and \(\Lambda\) are dimensionless. The last column is the fractional change \(X_{\rm reg}/X_{\rm bare}-1\). Regularized values use the EOS table with 256 points per decade. The smallest \(\Lambda\) offsets are comparable to the sensitivity to this table.}
\label{tab:bare-gr}
\centering
\setlength{\tabcolsep}{6pt}
\begin{tabular}{llrrr}
\hline
$M/M_\odot$ & $X$ & Bare GR & Regularized GR & $X_{\rm reg}/X_{\rm bare}-1$ \\
\hline
$0.6$ & $R\ [{\rm km}]$ & $8.837\times10^{3}$ & $8.878\times10^{3}$ & $4.667\times10^{-3}$ \\
$0.6$ & $k_2$ & $1.180\times10^{-1}$ & $1.153\times10^{-1}$ & $-2.301\times10^{-2}$ \\
$0.6$ & $\Lambda$ & $7.767\times10^{18}$ & $7.767\times10^{18}$ & $2.1\times10^{-7}$ \\
$1.3$ & $R\ [{\rm km}]$ & $3.185\times10^{3}$ & $3.188\times10^{3}$ & $7.727\times10^{-4}$ \\
$1.3$ & $k_2$ & $5.425\times10^{-2}$ & $5.404\times10^{-2}$ & $-3.855\times10^{-3}$ \\
$1.3$ & $\Lambda$ & $4.550\times10^{14}$ & $4.550\times10^{14}$ & $3.4\times10^{-10}$ \\
\hline
\end{tabular}
\end{table}
\endgroup

\flushbottom
\twocolumngrid

\bibliographystyle{apsrev4-2}
\bibliography{references}

\end{document}